\documentclass[aps,prx,superscriptaddress,twocolumn,10pt]{revtex4-2}
\usepackage{physics}
\usepackage{amsmath,amssymb}
\usepackage{braket}
\usepackage{graphicx}
\usepackage{xcolor}
\usepackage{verbatim}
\usepackage{color}
\usepackage{tabularx}
\usepackage{booktabs}
\usepackage{blindtext}
\usepackage{epstopdf}
\usepackage{nicefrac}
\usepackage{mathrsfs}
\usepackage{orcidlink}
\renewcommand{\vec}{\mathbf}

\makeatletter
\def\convertto#1#2{\strip@pt\dimexpr #2*65536/\number\dimexpr 1#1}
\makeatother

\newcommand{\vc}[1]{\boldsymbol{\mathrm{#1}}}
\newcommand{\mt}[1]{\mathrm{#1}}
\begin{document}
\title{Interferometric Braiding of Anyons in Chern Insulators}
\author{F.~A.~Palm\orcidlink{0000-0001-5774-5546}}
\affiliation{Department of Physics and Arnold Sommerfeld Center for Theoretical Physics (ASC), Ludwig-Maximilians-Universit\"at M\"unchen, Theresienstr. 37, D-80333 M\"unchen, Germany}
\affiliation{Munich Center for Quantum Science and Technology (MCQST), Schellingstr. 4, D-80799 M\"unchen, Germany}
\affiliation{CENOLI, Universit\'e Libre de Bruxelles, CP 231, Campus Plaine, B-1050 Brussels, Belgium}
\author{N.~Mostaan\orcidlink{0000-0002-9573-7608}}
\affiliation{Department of Physics and Arnold Sommerfeld Center for Theoretical Physics (ASC), Ludwig-Maximilians-Universit\"at M\"unchen, Theresienstr. 37, D-80333 M\"unchen, Germany}
\affiliation{Munich Center for Quantum Science and Technology (MCQST), Schellingstr. 4, D-80799 M\"unchen, Germany}
\affiliation{CENOLI, Universit\'e Libre de Bruxelles, CP 231, Campus Plaine, B-1050 Brussels, Belgium}
\author{N.~Goldman}
\affiliation{CENOLI, Universit\'e Libre de Bruxelles, CP 231, Campus Plaine, B-1050 Brussels, Belgium}
\affiliation{Laboratoire Kastler Brossel, Coll\`ege de France, CNRS, ENS-Universit\'e PSL, Sorbonne Universit\'e, 11 Place Marcelin Berthelot, 75005 Paris, France}
\affiliation{International Solvay Institutes, 1050 Brussels, Belgium}
\author{F.~Grusdt\orcidlink{0000-0003-3531-8089}}
\affiliation{Department of Physics and Arnold Sommerfeld Center for Theoretical Physics (ASC), Ludwig-Maximilians-Universit\"at M\"unchen, Theresienstr. 37, D-80333 M\"unchen, Germany}
\affiliation{Munich Center for Quantum Science and Technology (MCQST), Schellingstr. 4, D-80799 M\"unchen, Germany}
\date{\today}
\begin{abstract}
    Coherent control and braiding of anyons remain central challenges in realizing topologically protected quantum operations.
    We propose a Ramsey interferometry protocol to directly access the geometric phases associated with anyons in fractional Chern insulators.
    Our approach employs impurities with individually addressable internal states that bind to the anyons, allowing their adiabatic motion and exchange under full spatial control.
    By combining Ramsey and spin-echo sequences using one and two impurities, the protocol gives independent access to the Aharonov–Bohm and exchange contributions to the total geometric phase, thereby providing an unambiguous probe of anyonic statistics.
    Our scheme can potentially be implemented in cold-atom quantum simulators as well as in van der Waals heterostructures.
    Complementary finite-size simulations in non-interacting Chern insulators quantify the system sizes required to faithfully extract geometric phases, highlighting the role of edge effects.
    Our results establish impurity-based interferometry as a feasible route toward direct anyon braiding experiments in quantum simulators and lay the groundwork for future explorations of non-Abelian braiding and topological quantum control.
\end{abstract}
\maketitle

\section{Introduction}
Recent years have witnessed an outburst of research on various aspects of topologically ordered states of matter.
Theoretical predictions of their exotic nature and promising applications in quantum computing have made their experimental realization, probing, and understanding long-standing goals in modern condensed matter physics~\cite{Nayak2008}.
The quest to construct topological qubits has ushered intense efforts to create and manipulate anyonic excitations, particularly in solid-state platforms and digital quantum processors.
Unequivocal signatures of such excitations are their fractional charge and statistics, which have been observed in solid-state experiments through various anyon interferometry schemes~\cite{kobayashi2021shot,jonckheere2023anyonic,ronen2021aharonov,ruelle2023comparing,glidic2023cross,feldman2021fractional,lin2021quantized}.
Although schemes based on quasiparticle interference in solid-state devices have proven effective in revealing the existence of anyons~\cite{bartolomei2020fractional,nakamura2020direct,lee2023partitioning}, direct spatiotemporal control over individual anyons still needs to be achieved.
On digital quantum processors, non-trivial topologically ordered states and gate-based braiding of their excitations have been realized only recently:
Such platforms offer coherent control over individual Abelian or non-Abelian anyons in the context of the toric code~\cite{satzinger2021realizing,Iqbal2024} and related surface codes~\cite{andersen2023observation}.
However, the same degree of control over other topologically ordered systems, such as fractional quantum Hall states and topological quantum spin liquids, and the associated textbook-type adiabatic braiding have yet to be realized. 

Meanwhile, quantum simulation of topological band structures -- for example with cold atoms in optical lattices or photonic systems -- using artificial gauge fields has seen substantial progress~\cite{cooper2019topological,Ozawa2019}, also in the strongly interacting regime~\cite{tai2017microscopy}.
The recent realizations of \mbox{$\nicefrac{1}{2}$-Laughlin} states of two bosons in an optical lattice~\cite{Leonard2023}, a twisted optical cavity~\cite{Clark2020}, a rotating microtrap~\cite{Lunt2025}, and a circuit QED system~\cite{Wang2024} offer the prospect of creating and controlling individual anyonic quasihole excitations of such fractional quantum Hall states~\cite{paredes20011}.

Here, we propose an experimentally feasible interferometric protocol to create anyons and directly extract their fractional charge and braiding statistics using a quantum impurity.
To this end, we propose to create a pair of anyons using local pinning potentials~\cite{Raciunas2018,Macaluso2020,Wang2022,Palm2024}.
From this state it is possible to extract the braiding angle by first binding spin-$\nicefrac{1}{2}$ impurities to the anyons~\cite{mostaan2025anyon,wagner2025sensing} and then applying a combined Ramsey interferometry and spin echo sequence.
An even simpler form of our protocol also allows for direct measurements of the Aharonov-Bohm phase, thus giving unbiased access to the pure exchange phase and to the anyons' fractional charge.

Below, we will first discuss our protocol and comment on its experimental feasibility.
In principle, our protocol is applicable to a broad class of systems, among them traditional solid-state platforms and state-of-the-art quantum simulators.
We furthermore perform numerical simulations of non-interacting Chern insulators, investigating the system sizes needed to unambiguously extract the Aharonov-Bohm and exchange phases.
We find that systems of at least a few hundred lattice sites are needed due to substantial edge effects.
While we expect this to be true for strongly interacting fractional Chern insulators as well, these system sizes are currently still out of reach for numerical simulations as well as experimental realizations in synthetic quantum materials.

\section{Impurity interferometry}
\begin{figure*}
    \centering
    \includegraphics{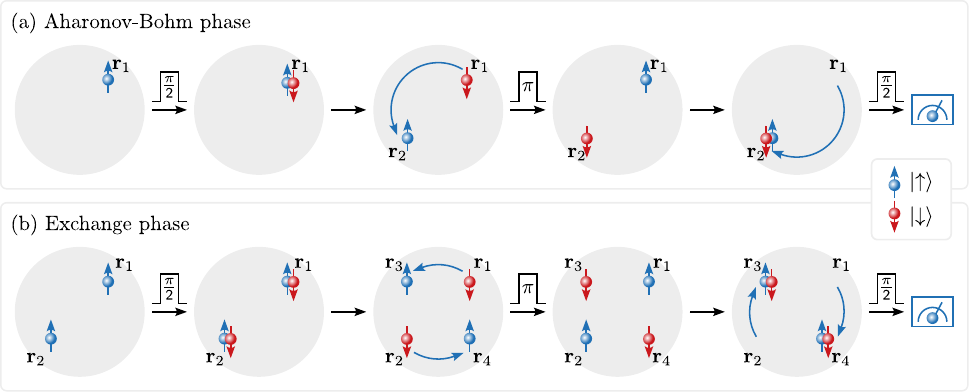}
    \caption{
        Interferometric sequences allowing for measurements of (a) the Aharonov-Bohm phase and (b) the exchange phase using impurities and state-dependent paths.
        Due to the combined Ramsey and spin echo sequence, the dynamical phase at the end of the sequence vanishes, resulting only in the measurement of the geometric contribution.
        Notably, for this sequence to work it is sufficient to bind $\ket{\uparrow}$ to the quasihole, whereas $\ket{\downarrow}$ does not need to move and can be completely decoupled from the many-body system.
    }
    \label{fig:SketchABPhase}
\end{figure*}
To directly measure geometric phases, we propose an interferometric protocol inspired by the approach in Ref.~\cite{Grusdt2016} for measuring the Chern number of a state containing anyons -- assumed here, for concreteness, to be quasiholes.
To this end, we suggest to immerse impurities with internal states $\ket{\uparrow}$ and $\ket{\downarrow}$ in the fractional quantum Hall liquid.
Our approach is general and can, in principle, be realized in a variety of physical platforms supporting fractional quantum Hall states. For clarity and specificity, however, we describe its implementation in the context of cold-atom quantum simulators.
%
%In this setting, we consider impurities with two internal states $\ket{\uparrow}$ and $\ket{\downarrow}$, immersed in a fractional quantum Hall liquid.
%
The impurities interact repulsively with the particles in the fractional quantum Hall liquid, with interaction strengths $v_{\uparrow(\downarrow)}$ which depend on the impurities' internal states and are proportional to the impurity-atom scattering lengths $a_{\uparrow(\downarrow)} \! > \! 0$. 
When the repulsive interaction between the impurity and the majority atoms is sufficiently strong for a particular internal state -- taken here to be \(\ket{\uparrow}\) -- an impurity localized at the position of a quasihole can effectively \textit{pin} the quasihole.
While an impurity in the \(\ket{\downarrow}\)-state could, in principle, also interact with the majority atoms -- and its interaction strength could potentially be tuned to match that of the \(\ket{\uparrow}\)-state -- a more experimentally realistic scenario is one in which only the \(\ket{\uparrow}\)-state interacts strongly.
In this case, the quasihole remains bound exclusively to the \(\ket{\uparrow}\)-impurity.
This scenario relies on the assumption that quasiholes possess a lifetime significantly longer than the duration of the interferometric sequence, such that they remain stable even in the absence of an external potential (e.g., from a \(\ket{\downarrow}\)-impurity or laser-induced trap).
If the impurity in the state $\ket{\uparrow}$ is now moved adiabatically, the quasihole remains bound to the impurity due to the strong repulsive interaction, thereby following its trajectory.
In this way, the position and motion of the quasihole can be controlled via the impurity, enabling the implementation of braiding operations.
For interferometric use, the $\ket{\downarrow}$-component of the impurity is not moved and thus not braided, thereby serving as the second arm of an interferometer, providing a reference phase relative to which the geometric phase is measured.

When two quasiholes are braided, the geometric phase acquired by the system consists of two contributions: an Aharonov-Bohm phase and an exchange phase.
The Aharonov-Bohm phase arises from the coupling of the quasihole density to the applied magnetic flux and is present even for a single quasihole, while the exchange phase results purely from the non-trivial statistics associated with the braiding of the quasiholes.
In the following, we analyze the role of each contribution in detail.

\subsection{One impurity: the Aharonov-Bohm phase}
We start our discussion by considering a single impurity immersed in the system, pinning a quasihole.
Moving the impurity along a closed path, the system's state acquires an Aharonov-Bohm phase, due to the coupling of the quasihole's density to the effective magnetic flux realizing the quantum Hall state.
The impurity is initially prepared in the $\ket{\uparrow}$-state and located at the position of the quasihole.
Applying a Ramsey $\nicefrac{\pi}{2}-$pulse and assuming spin-dependent trapping potentials $g_{\uparrow}(\vc{r})\!\neq\! g_{\downarrow}(\vc{r})$ seen by the impurity allows to independently control the position of the $\ket{\uparrow}$- and $\ket{\downarrow}$-component of the impurity.
By moving the $\ket{\uparrow}$-impurity along a closed loop while keeping the $\ket{\downarrow}$-impurity fixed, see Fig.~\ref{fig:SketchABPhase}(a), we will show now that the geometric Aharonov-Bohm phase can be extracted directly using Ramsey interferometry.
To eliminate the dynamical phase accumulated during the interferometric sequence, a spin-echo protocol is employed.
Specifically, the closed path traversed by the impurity is divided into two equivalent segments.
To ensure that the dynamical phases acquired along each segment are equal, the two paths are designed to be mirror-symmetric with respect to the line connecting the initial and final positions of the impurity (see Fig.~\ref{fig:SketchABPhase}).
In this configuration, applying a $\pi$-pulse midway through the sequence completes the spin-echo and effectively cancels the total dynamical phase.
We will further develop the details of the sequence below.

The sequence starts with the impurity in the state $\ket{\uparrow}$ trapped at the position of the quasihole $\vc{r}_1$, such that the initial state is
\begin{equation}\label{eq:psi_i:1ptcl}
    \ket{\Psi_{\rm init}}=\ket{\vc{r}_1 \uparrow} \ket{\mathrm{qh}}_{\mathrm{QH}} .
\end{equation}
Here, $\ket{\mathrm{qh}}_{\rm QH}$ denotes the initial quantum Hall state with a quasihole.
Applying a $\nicefrac{\pi}{2}$-pulse to the impurity changes its spin state from $\ket{\uparrow}$ to \mbox{$\ket{+}\!=\!\big(\ket{\uparrow}+\ket{\downarrow}\big)/\sqrt{2}$}, thus changing the state of the system to $\ket{\vc{r}_1 +} \ket{\mathrm{qh}}_{\mathrm{QH}}$.
Next, the potential $g_{\uparrow}(\vc{r})$ is used to move the $\ket{\uparrow}$-state along the first half of the closed path discussed above, while the potential $g_{\downarrow}(\vc{r})$ keeps the $\ket{\downarrow}$-state at the initial position, see Fig.~\ref{fig:SketchABPhase}(a).
After one half of the interferometry sequence, a spin echo $\pi-$pulse implements a spin flip for the impurity, before the $\ket{\uparrow}$-impurity is moved from $\vc{r}_1$ to $\vc{r}_2$, to the position of the $\ket{\downarrow}$-impurity.
A second $\nicefrac{\pi}{2}$-pulse with a control phase $\phi_c$ closes the Ramsey sequence and a joint measurement of the impurities' spin and position results in the state $\ket{\vc{r}_2\uparrow}$ with probability
\begin{equation}
    p_{\uparrow}=|\left(1+\mathrm{exp}[i(\varphi_{\mathrm{geo}}-\phi_c)]\right)/2|^2=\mathrm{cos}^2\left(\frac{\varphi_{\rm geo}-\phi_c}{2}\right),
    \label{Eq:RamseyFormula:1ptcl}
\end{equation}
where $\varphi_{\rm geo}=\varphi_{\rm AB}$ is the geometric Aharonov-Bohm phase acquired by the quasihole moved along a closed path.

In summary, the measurement of $\ket{\vc{r}_2 \uparrow}$ after moving the impurity along a symmetry-preserving, closed path gives direct access to the Aharonov-Bohm phase $\varphi_{\rm AB}$.
For additional details regarding the evolution of the state throughout the sequence, we refer the reader to the supplemental material~\cite{supp}.

\subsection{Two impurities: the exchange phase}
Next, we turn to the case of two impurities, which allows for measurements of the exchange phase and therefore also gives insight into the braiding phase and the statistics of the quasiholes.
Similar to the single impurity case, the impurities are initially prepared in their $\ket{\uparrow}$-state, bound to the quasiholes at positions $\vec{r}_1$ and $\vec{r}_2$,
\begin{equation}\label{eq:psi_i}
    \ket{\Psi_{\rm init}}=\ket{\vc{r}_1 \uparrow,\vc{r}_2\uparrow} \ket{\mathrm{2-qh}}_{\mathrm{QH}} .
\end{equation}
Again, after applying a global Ramsey $\nicefrac{\pi}{2}-$pulse on both impurities, spin-dependent potentials allow to independently control the position of the $\ket{\uparrow}$- and $\ket{\downarrow}$-component of the impurities.
Modifying our previous protocol to incorporate an exchange of the quasiholes along one path of the interferometer, see Fig.~\ref{fig:SketchABPhase}(b), the exchange phase is extracted directly using a combination of Ramsey interferometry and a spin echo sequence.

To conclude, the measurement of $\ket{\vc{r}_3 \uparrow,\vc{r}_4 \uparrow}$ gives direct access to the geometric phase $\varphi_{\rm geo}$ via
\begin{equation}
    p_{\uparrow \uparrow}=|\left(1+\mathrm{exp}[i(\varphi_{\mathrm{geo}}-2\phi_c)]\right)/4|^2=\frac{1}{4}\,\mathrm{cos}^2\left(\frac{\varphi_{\rm geo}-2\phi_c}{2}\right).
    \label{Eq:RamseyFormula}
\end{equation}
Note that in this case both the (single impurity) Aharonov-Bohm phase $\varphi_{\rm AB}$ and the (two impurity) exchange phase $\varphi_{\rm exc}$ contribute to the geometric phase, $\varphi_{\rm geo} \!=\! \varphi_{\rm exc} + \varphi_{\rm AB}$.
Combining the full geometric phase obtained with two impurities with the single-impurity Aharonov-Bohm measurement allows one to isolate the exchange part.
For additional details, we refer the reader to the supplemental material~\cite{supp}.

\subsubsection*{The non-Abelian case}
We would like to emphasize that similar protocols can also be employed to study non-Abelian anyons, for example in the Moore-Read Pfaffian state~\cite{Moore1991}.
For such non-Abelian states, the low-energy manifold in the presence of some anyons consists of degenerate states $\ket{\psi_{a}}_{\rm QH}$, $a\!=\!1, \hdots, M$.
To generalize the Abelian geometric phase $\exp[i\varphi_{\rm geo}]$ to the non-Abelian case, we use the Wilson loop operator
\begin{equation}
    \hat{W} = \exp\left[i \oint \mathrm{d}\vec{\lambda} \cdot \vec{\mathcal{A}}(\vec{\lambda})\right],
\end{equation}
where $\vec{\lambda}$ parametrizes a closed loop in parameter space -- e.g. a braiding path -- and $\vec{\mathcal{A}}_{\mu}^{ab} \!=\! \braket{\psi_{a}|i\partial_{\lambda_{\mu}}|\psi_{b}}_{\rm QH}$ denotes the non-Abelian Berry connection.
We envision that all matrix elements of the Wilson loop operator $\hat{W}$, defined in the manifold of degenerate states, can be systematically mapped out.
In particular, the amplitude of the interferometric signal is proportional to $\braket{\psi|\hat{W}|\psi}_{\rm QH}$ via
\begin{equation}\label{eq:W}  
    p_{\uparrow\uparrow} = \frac{1}{8}\,\left(1+|\braket{\psi|\hat{W}|\psi}| \cos(\varphi^{\psi}-2\phi_c)\right) \, ,
\end{equation}
where $\varphi^{\psi} = \arg\braket{\psi|\hat{W}|\psi}$ and $\ket{\psi}$ is a combination of the basis states $\ket{\psi_{a}}$ in the degenerate manifold.
In practice, this could be achieved via internal (braiding) operations within the degenerate manifold.
Therefore, our interferometric protocol allows for a complete classification of the braiding rules of a given non-Abelian state.

\subsection{Experimental implementation}
In cold atom quantum simulators, our impurity interferometry scheme could be realized by preparing a $\nu=\nicefrac{1}{2}$-Laughlin state of $^{87}$Rb atoms in the $\ket{F=1, m_{F}=-1}$ state \cite{Leonard2023} and using two $^{87}$Sr atoms as impurities.
The availability of a $^{87}$Rb-$^{87}$Sr Feshbach resonance at $B=521.5(4)\,G$ \cite{barbe2018observation} allows to achieve strong impurity-majority interactions.
The clock states of $^{87}$Sr (the ground $^{1}\mathrm{S}{0}$ and excited $^{3}\mathrm{P}{0}$ states) enable state-dependent control of the impurity position \cite{heinz2020state}, while the interaction strength between $^{87}$Rb atoms and the $^{1}\mathrm{S}_{0}$ state of $^{87}$Sr can be tuned via the aforementioned Feshbach resonance.
Our protocol might also be applicable to experiments with fermionic $^6$Li~\cite{Lunt2025}, where additional hyperfine states could be used to realize the impurities.

Furthermore, our protocol is not limited to cold atomic systems, but also holds promise for implementation in solid-state platforms based on van der Waals heterostructures. 
In these systems, stable electronic fractional quantum Hall states can be realized within a single layer, while adjacent optically active layers -- such as those based on transition metal dichalcogenides (TMDs) -- can serve as probe layers.
In such configurations, tightly bound excitons can be optically injected via resonant excitation in the TMD layers. Due to their sharp and stable resonances, these excitons can act as sensitive optical probes of the fractional quantum Hall system~\cite{popert2022optical,cui2024interlayer,gao2025probing}.
In such settings, a TMD heterostructure is placed in proximity to a quantum Hall system -- for example, a fractional quantum Hall state in a graphene monolayer or a fractional Chern insulator in a twisted \(\mathrm{MoT}_2\) bilayer.
In bilayer TMDs, interlayer excitons can interact with electrons either repulsively or attractively, depending on the layer configuration of the electron and hole within the exciton.
For sufficiently strong repulsive interactions, excitons can bind to quasiholes in the quantum Hall system~\cite{mostaan2025anyon,wagner2025sensing}.
The ability to generate strongly localized excitons—with spatial extents on the order of \(\sim 1~\mathrm{nm}\)—combined with the development of optical quantum twist microscopes (QTMs), where an exciton can be localized at the microscope tip, may enable precise spatial and motional control of exciton positioning~\cite{mostaan2025anyon,wagner2025sensing}.
Furthermore, the internal degrees of freedom of excitons can be manipulated using polarized light, allowing for the creation of superpositions of valley states and enabling the implementation of the same impurity interferometry protocol described below.

Key challenges in this solid-state context include the relatively short exciton lifetimes, which can limit the extent of mechanical motion before recombination, and the difficulty in maintaining long-lived coherence in excitonic states.
These limitations are mitigated in cold atomic systems, where ultra long coherence times of clock states and greater control over both internal states and interactions are readily achievable.
However, this advantage comes at the cost of substantially smaller system sizes currently available in experiments.

\section{Numerical finite-size studies}
While solid state platforms can be considered to directly operate in the thermodynamic limit, many quantum simulation platforms are still limited in system size.
To study the role of finite-size effects in such systems, we perform numerical simulations of finite systems.
We provide numerical evidence for the applicability of our adiabatic protocol to directly extract the Aharonov-Bohm and exchange phases, given that the path of the pinning potentials is well-separated from the edge of the system.
We show for a paradigmatic non-interacting Chern insulator that the system sizes needed to extract a clean signal are on the order of a few hundred lattice sites.
Based on estimates of the characteristic length scales, we expect similar sizes to be needed for more complex fractional Chern insulators.
We furthermore show how a local magnetic flux can be used to obtain the (quasi)hole charge from Aharonov-Bohm phase measurements directly.

\subsection{Model and methods}
Inspired by the recent progress in quantum simulators~\cite{Leonard2023,Wang2024}, we study the Hofstadter-Hubbard model on a square lattice.
In Landau gauge the Hamiltonian reads
\begin{widetext}
\begin{equation}
        \hat{\mathcal{H}} = -J \sum_{x,y} \left(\hat{a}^{\dagger}_{x+1,y}\hat{a}_{x,y}^{\vphantom\dagger} + \mathrm{e}^{i 2\pi \left(\alpha + \delta\alpha(x,y)\right) x}\hat{a}^{\dagger}_{x,y+1}\hat{a}_{x,y}^{\vphantom\dagger} + \mathrm{H.c.}\right)
        +\frac{U}{2}\sum_{x,y} \hat{n}_{x,y}\left(\hat{n}_{x,y}-1\right)
    \label{eq:HofstadterHubbardModel}
\end{equation}
\end{widetext}
where $\hat{a}^{(\dagger)}_{x,y}$ annihilates (creates) a particle at site $(x,y)$, $\hat{n}_{x,y} \!=\! \hat{a}_{x,y}^{\dagger}\hat{a}_{x,y}^{\vphantom\dagger}$ is the particle number operator, $(x_0, y_0)$ denotes the center of the lattice, and $\alpha$ is the background flux per plaquette.
In addition, we introduce an additional flux $\delta\alpha(x,y)$ through some plaquettes, as specified below.
In our numerical simulations we restrict ourselves to non-interacting fermions, thus setting $U/J=0$.
All our simulations are performed for open boundary conditions, relevant also for experiments.
Anyon braiding on a small torus showed the conceptual applicability of a similar protocol under idealized conditions~\cite{Kapit2012}.

To trap and move (quasi-)particles and holes, we introduce Gaussian pinning potentials of the form
\begin{equation}
    \hat{V}_{\rm pin} = V_{\rm pin} \sum_{x,y} \mathrm{e}^{-((x-X)^2 + (y-Y)^2)/2\sigma^2} \hat{n}_{x,y},
\end{equation}
where $\vec{R} \!=\! (X, Y)$ is the center of the pinning potential and $\sigma$ is its width.
In our simulations, we vary the center $\vec{R}$ of the pinning potential along a closed loop $\mathcal{C}$, which we discretize into $N_{\rm steps}$ evenly spaced steps with center position $\vec{R}_{j}$, $j\!=\!0, \hdots, N_{\rm steps}$, where we explicitly include the last point such that $\vec{R}_{0} \!=\! \vec{R}_{N_{\rm steps}}$.
In each step, we calculate the ground state $\ket{\psi_{j}}$ of the system, hence simulating a perfectly adiabatic protocol.
Afterwards, we use this set of ground states to determine the geometric phase acquired along the path $\mathcal{C}$,
\begin{equation}
    \varphi_{\rm geo} = \mathrm{Im} \log\left( \prod_{j=0}^{N_{\rm steps}} \braket{\psi_{j+1}|\psi_{j}}\right),
    \label{Eq:GeometricPhaseDiscretizedPath}
\end{equation}
where we included the factor $\braket{\psi_0|\psi_{N_{\rm steps}}}$ to remove the overall gauge-redundancy of the expression, see Ref.~\cite{Resta1994}.

\subsection{Single particle in a magnetic field}
\label{sec:Benchmark}

\subsubsection{Encircling a small flux region}
We start our simulations with the particularly simple case of a single particle in a magnetic field, to confirm the applicability of our numerical tools.
We turn off the background magnetic flux, $\alpha\!=\!0$, and only introduce a local magnetic flux $\delta\Phi = 4\times \delta\alpha$ through the central four plaquettes.
We consider an attractive pinning potential of strength $V_{\rm pin}/J \!=\! -5$ and width $\sigma/a\!=\!1$ which is moved along a circular path of radius $R$.
In our simulations, we systematically vary $R$ and split the path into $N_{\rm steps}\!=\!40$ steps to adiabatically follow the ground state.
To allow for a wide range of possible radii $R$ and to avoid edge effects, we consider a system of size $15\times15$.

The predicted Aharonov-Bohm phase for a particle of charge $q^{\star}$ moved along such a path is
\begin{equation}
    \varphi_{\rm AB}^{\rm (pred)}(\delta\alpha) = 2\pi\delta\alpha \min\left(4, \pi \left(\nicefrac{R}{a}\right)^2\right) \times q^{\star},
    \label{eq:ABPhase:singleParticle}
\end{equation}
where the $\min$-expression takes into account that the path only partially encircles the flux region for $R/a\!<\!2 $.

\begin{figure}
    \centering
    \includegraphics{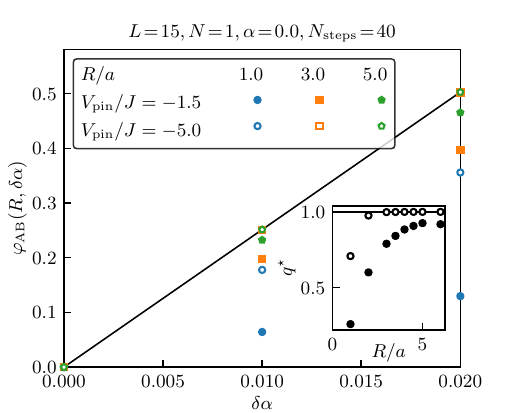}
    \caption{
        Aharonov-Bohm phase obtained from Eq.~\eqref{Eq:GeometricPhaseDiscretizedPath} for a single particle encircling a flux $\delta\alpha$ through the central four plaquettes along a circular path of radius $R/a$ for $V_{\rm pin}/J=-1.5$ (full symbols) and $-5.0$ (open symbols), respectively.
        For $R/a\!=\!1$ the path does not encircle the local flux completely, resulting in substantial deviations from the expected behavior.
        For radii $R/a \!\gtrsim\! 2$ the entire flux is encircled, so that we expect a scaling \mbox{$\varphi_{\rm AB} \!\propto\! 2\pi \times 4\delta\alpha \times q^{\star}$} according to Eq.~\eqref{eq:ABPhase:singleParticle}, with $q^{\star}$ the charge of the pinned object.
        The solid black line indicates this scaling for $q^{\star}\!=\!1$.
        We find that for a sufficiently strong pinning potential the numerically extracted Aharonov-Bohm matches this prediction as well as the expected $q^{\star}\!=\!1$ (see inset).
        Here, $q^{\star}$ is obtained using a fit of $\varphi_{\rm AB}^{(\rm pred)}(\delta\alpha)$ to the numerical data.
    }
    \label{fig:ABphaseSingleParticle}
\end{figure}

We find that the Aharonov-Bohm phase extracted from Eq.~\eqref{Eq:GeometricPhaseDiscretizedPath} along this path agrees well with the theoretical prediction, as long as the particle does not overlap with the flux region, see Fig.~\ref{fig:ABphaseSingleParticle}.
By performing a fit of $\varphi_{\rm AB}^{(\rm pred)}(\delta\alpha)$ in Eq.~\eqref{eq:ABPhase:singleParticle} to the numerically obtained Aharonov-Bohm phases, we can determine the charge $q^{\star}$ of the object pinned by the potential.
We find that the extracted charge is consistent with $q^{\star} \!=\! 1$ for paths staying sufficiently far away from the  flux region, $R/a\!\gtrsim\! 2$, see the inset of Fig.~\ref{fig:ABphaseSingleParticle}.

Reducing the strength of the pinning potential ($V_{\rm pin}/J=-1.5$) results in quantitatively less accurate results, while preserving the qualitative features.

\subsubsection{Homogeneous magnetic background field}
Next, we perform similar simulations in the presence of a homogenoeus magnetic background flux $\alpha\!=\!0.2$ per plaquette, thus creating topologically non-trivial Chern bands.
To adiabatically follow the ground state and avoid Chern band mixing, we choose a pinning potential of strength $V_{\rm pin}/J \!=\! -1.0$, smaller than the band gap $\Delta/J \approx 1.5$ above the lowest band, and width $\sigma/a \!=\! 1.0$.

We thread an additional flux $\delta\Phi \!\leq\! 0.08$ through the central four plaquettes to probe the dependence of the geometric phase on such a flux defect.
To fully encircle the region of additional flux, we choose radii $R/a \!>\! 2$.

\begin{figure}
    \centering
    \includegraphics{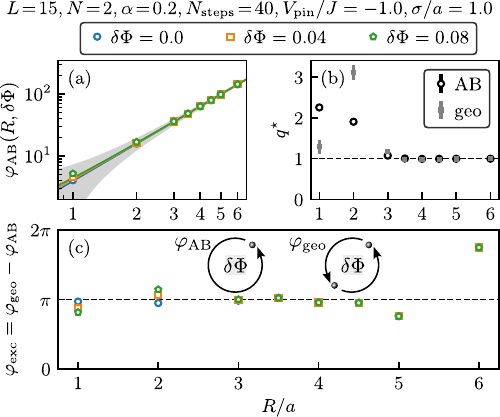}
    \caption{
        (a) Aharonov-Bohm phase for a single particle in a magnetic field of flux $\alpha\!=\!0.2$ per plaquette, encircling an additional flux $\delta\Phi$ through the central four plaquettes along a circular path of radius $R/a$.
        Already for a relatively weak pinning potential ($V_{\rm pin}/J\!=\!-1,~ \sigma/a\!=\!1$) and $R/a\!\gtrsim\! 3$ the numerically extracted Aharonov-Bohm phase (symbols) matches the prediction (lines).
        The solid lines indicate the expected scaling for the respective $\delta\Phi$, whereas the gray shaded area indicates a band of width $2\pi$ around the expected value for $\delta\alpha\!=\!0$.
        (b) Pinned charge $q^{\star}$ as extracted from the Aharonov-Bohm (circles) and full geometric (squares) phases.
        For $R/a\gtrsim3$ the pinned charge agrees well with the expected $q^{\star}\!=\!1$.
        Here, $q^{\star}$ is obtained using a fit of $\varphi_{\rm AB/geo}^{(\rm pred)}(\delta\Phi)$ to the numerical data, respectively.
        Note, that measurements of $\varphi_{\rm AB}$ involve only one particle, whereas measurements of the full geometric phase $\varphi_{\rm geo}$ require to manipulate two particles.
        (c) Extracted exchange phase for two fermions.
        For $3 \!\lesssim\! R/a \!\lesssim\! 4.5$, the phase is consistent with the fermionic $\varphi_{\rm exc}\!=\!\pi$, whereas for smaller radii the effect of not fully encircling the additional flux $\delta\Phi$ is visible for $\delta\Phi \!\neq\! 0$.
        For larger radii, edge effects become substantial and lead to deviations from the expected exchange phase.
        The insets show the different paths taken by the pinning potentials to determine the Aharonov-Bohm (left) and full geometric (right) phases, respectively.
    }
    \label{fig:ABphasesSingleParticle:BackgroundFlux}
\end{figure}

The expected Aharonov-Bohm phase for such a system is
\begin{equation}
    \varphi_{\rm AB}^{\rm(pred)}(\delta\Phi) = 2\pi q^{\star} \left(\alpha \times \pi \left(\nicefrac{R}{a}\right)^2 + \delta\Phi\right).
    \label{eq:FullABPhase}
\end{equation}
Simulating the system for various values of $R/a$ and $\delta\alpha$, we confirm the scaling $\varphi_{\rm AB} \!\sim\! R^2$, see Fig.~\ref{fig:ABphasesSingleParticle:BackgroundFlux}(a).
We also find a linear dependence of $\varphi_{\rm AB}$ on $\delta\Phi$, which we use to extract the charge $q^{\star}$ of the pinned object, which is consistent with $q^{\star} \!=\! 1$ for sufficiently large $R/a \!\gtrsim\! 3$, see Fig.~\ref{fig:ABphasesSingleParticle:BackgroundFlux}(b).

We remark that for a finite external flux $\alpha \!\neq\!0$ the pinning potential needed to trap the particle is substantially weaker than for the case without external magnetic field.
We attribute this effect to the formation of flat Chern bands, which tend to localize the particles and hence increase the relative effect of already weak pinning potentials.

\subsection{Exchanging two fermions}
We perform a minimal numerical ``braiding'' experiment by exchanging two fermions in the presence of a magnetic field.
In this case, the full geometric phase is given by
\begin{equation}\label{eq:geometricPhase}
    \varphi_{\rm geo} = \varphi_{\rm AB} + \varphi_{\rm exc},
\end{equation}
where the predicted Aharonov-Bohm phase $\varphi_{\rm geo}$ is given by the expression in Eq.~\eqref{eq:FullABPhase} and the exchange phase $\varphi_{\rm exc} \!=\! \pi$ for fermions.
Experimentally, the Aharonov-Bohm phase can be determined by moving a single particle along a closed path, whereas the full geometric phase is obtained from the exchange of two particles along the first half of this path, see also Fig.~\ref{fig:SketchABPhase}.
Thereby, the pure exchange phase can be extracted as \mbox{$\varphi_{\rm exc} \!=\! \varphi_{\rm geo} - \varphi_{\rm AB}$}.
We emphasize that while the detection of the Aharonov-Bohm phase $\varphi_{\rm AB}$ is a purely single-particle measurement, the measurement of the full geometric phase $\varphi_{\rm geo}$ involves two particles.

Guided by our earlier results, we consider a system of $15\times15$ sites and vary the radius $R$ of the exchange path from the center of the system.
At the same time, we insert a small flux $\delta\Phi \!=\! 0.00 \hdots 0.08$ through the central four plaquettes, thus allowing us to extract the pinned charge $q^{\star}$ from a linear fit of $\varphi_{\rm AB}(\delta\Phi)$.
As before, we fix the parameters of the pinning potential to \mbox{$V_{\rm pin}/J \!=\! -1.0,~ \sigma/a \!=\! 1.0$}.
We find that for radii $3 \!\lesssim\! R/a \!\lesssim\! 4.5$ the extracted exchange phase is consistent with $\varphi_{\rm exc} \!=\! \pi$, which for $\delta\Phi\!=\!0.0$ is true even for smaller radii, see Fig.~\ref{fig:ABphasesSingleParticle:BackgroundFlux}(c).
For even larger radii, $R/a\!\gtrsim\!4.5$, edge effects become substantial and lead to deviations from the expected exchange phase.
Furthermore, we find that for $R/a \!\gtrsim\! 3$ the charge extracted from $\varphi_{\rm geo}(\delta\Phi)$ is consistent with $q^{\star} \!=\! 1$, see Fig.~\ref{fig:ABphasesSingleParticle:BackgroundFlux}(b).

We conclude that our protocol is suitable to extract the exchange phase given an exchange path which avoids both the $\delta\Phi$-flux region in the center and the edge of the system.

\subsection{Non-interacting Chern insulator}
\begin{figure}
    \centering
    \includegraphics{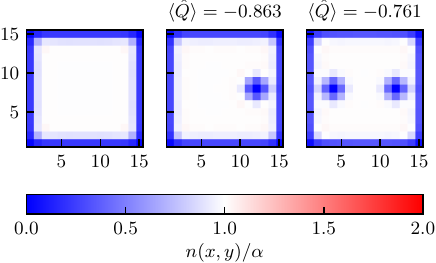}
    \caption{
        Density profiles for a system of $N\!=\!35$ particles in a magnetic field of flux $\alpha\!=\!0.2$ per plaquette subject to no (left), one (middle) and two (right) pinning potentials (\mbox{$V_{\rm pin}/J \!=\! 1.5,~ \sigma/a\!=\!1.0$}).
        The ground state without any pinning potentials realizes a Chern insulator in the lowest Hofstadter band.
    }
    \label{fig:ChernInsulator:Density}
\end{figure}

Having confirmed the validity of our approach for the minimal case of two fermions, we now continue by probing the simplest non-trivial many-body state.
In particular, we consider a non-interacting Chern insulator of $N\!=\!35$ fermions on $15\times15$ sites pierced by a magnetic flux $\alpha\!=\!0.2$ per plaquette.
For these parameters, the bulk of the systems exhibits a Chern insulating state at $\nu\!=\!1$.
We again introduce an additional local flux $\delta\Phi$ through the central four plaquettes of the system.
We choose pinning potentials of strength $V_{\rm pin}/J \!=\! 1.5$ and width $\sigma/a\!=\!1.0$, strong enough to pin one and two holes on top of the $\nu\!=\!1$ Chern insulator, see Fig.~\ref{fig:ChernInsulator:Density}.
We define the charge operator
\begin{equation}
    \hat{Q} = \sum_{\vec{r}} \mathrm{e}^{-(\vec{r}-\vec{R})^2/\xi^2} \left(\hat{n}^{\vphantom{(0)}}_{\vec{r}} - n^{(0)}_{\vec{r}}\right),
    \label{eq:ChargeOperator}
\end{equation}
where $n^{(0)}_{\vec{r}}$ is the density in the absence of the pinning potentials, $\vec{R}$ is the center site of one pinning potential, and $\xi\!=\!2a$ is the width of a Gaussian envelope function chosen to be larger than the magnetic length $\ell_B \!=\! \nicefrac{a}{\sqrt{2\pi\alpha}} \!\approx\! 1.25 a$ but smaller than the distance to the edge.
By evaluating the expectation value $\braket{\hat{Q}}$, we obtain an estimate of the pinned charge based on ground state properties.
We find that the pinned charge $\braket{\hat{Q}}=-0.863$ for one pinning potential ($\braket{\hat{Q}}=-0.761$ for two potentials) is roughly consistent with a pinned hole of (negative) unit charge, while systematically underestimating the pinned (negative) charge.

\begin{figure}
    \centering
    \includegraphics{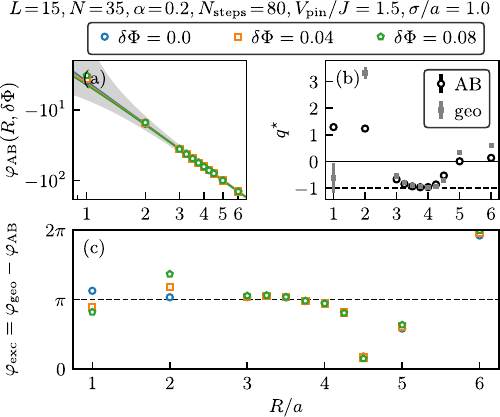}
    \caption{
        (a) Aharonov-Bohm phase for a single pinning potential ($V_{\rm pin}/J\!=\!1.5,~ \sigma/a\!=\!1$) moved along a circular path of radius $R/a$ in a system of $N\!=\!35$ particles on $15\times15$ sites with a flux $\alpha\!=\!0.2$ per plaquette, realizing a Chern insulator in the lowest Hofstadter band.
        The solid lines indicate the expected scaling for the respective $\delta\alpha$, whereas the gray shaded area indicates a band of width $2\pi$ around the expected value for $\delta\Phi\!=\!0$.
        (b) Pinned charge $q^{\star}$ as extracted from the Aharonov-Bohm (circles) and full geometric (squares) phases.
        For $3 \!\lesssim\! R/a \!\lesssim\! 4$ the pinned charge agrees well with the expected $q^{\star}\!=\!-1$.
        Here, $q^{\star}$ is obtained using a fit of $\varphi_{\rm AB/geo}^{(\rm pred)}(\delta\Phi)$ to the numerical data, respectively.
        (c) Extracted exchange phase for two pinning potentials.
        For $3 \!\lesssim\! R/a \!\lesssim\! 4$, the phase is consistent with the expected $\varphi_{\rm exc}\!=\!\pi$, whereas for smaller radii the effect of not fully encircling the additional flux $\delta\Phi$ is visible for $\delta\Phi \!\neq\! 0$.
        For larger radii, edge effects become substantial and lead to deviations from the expected exchange phase.
    }
    \label{fig:ChernInsulator:ABPhase:N35}
\end{figure}

Next, we move a single hole around a closed path split into $N_{\rm steps}\!=\!80$ segments.
We extract the Aharonov-Bohm phase $\varphi_{\rm AB}(R, \delta\Phi)$ in the presence of the Chern insulating bulk and find that for $R/a \!\approx\! 4$ the charge extracted from the Aharonov-Bohm phase is consistent with $q^{\star}\!=\!-1$, see Fig.~\ref{fig:ChernInsulator:ABPhase:N35}(a,b).
For larger radii edge effects become sizable.

Finally, we exchange two pinning potentials along a path of $N_{\rm steps}\!=\!40$ steps.
We find that for radii \mbox{$3 \!\lesssim\! R/a \!\lesssim\! 4$} the extracted exchange phase is consistent with $\varphi_{\rm exc} \!=\! \pi$, which
for $\delta\Phi\!=\!0.0$ is true even for smaller radii, see Fig.~\ref{fig:ChernInsulator:ABPhase:N35}(c).
Furthermore, we find that in this regime the charge extracted from $\varphi_{\rm geo}(\delta\Phi)$ is consistent with $q^{\star} \!=\! -1$, see Fig.~\ref{fig:ChernInsulator:ABPhase:N35}(b).
Again, for large radii ($R/a \!\gtrsim\! 4.5$) edge effects become sizable and the exchange phase cannot be extracted in a meaningful manner.
Similar simulations with an additional projection to the lowest Chern band give essentially unchanged results~\cite{supp}.

\begin{figure}
    \centering
    \includegraphics{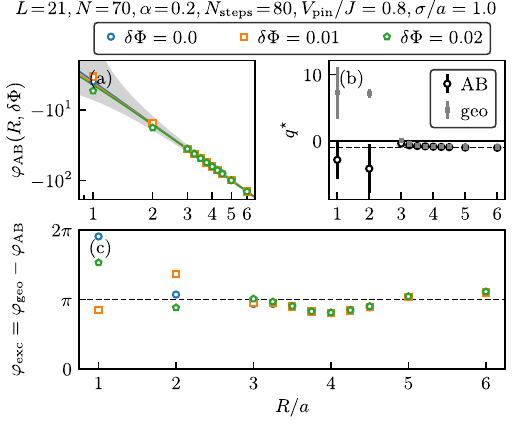}
    \caption{
        Same as Fig.~\ref{fig:ChernInsulator:ABPhase:N35}, but for a system of $N\!=\!70$ particles on $21\times21$ sites with a pinning potential of strength $V_{\rm pin}/J\!=\!0.8$.
        We find that for circular paths of radius $R/a \gtrsim 3$ all observables agree well with their expected values and the effect of the edge of the system is significantly less pronounced than in the smaller system.
    }
    \label{fig:ChernInsulator:ABPhase:N70}
\end{figure}

To study the role of the edge further, we perform a similar analysis for $N\!=\!70$ particles on $21\times21$ sites, see Fig.~\ref{fig:ChernInsulator:ABPhase:N70}.
Again, the Aharonov-Bohm phase is calculated for circular paths of varying radius, discretized into $N_{\rm steps}\!=\!80$ steps.
We apply pinning potentials of strength $V_{\rm pin}/J \!=\! 0.8$ and width $\sigma/a \!=\! 1.0$.
The pinning potential was chosen such that the same parameters allow for pinning (exactly) one and two holes.

We find that using a larger system substantially increases the range of radii for which the charge extracted from the Aharonov-Bohm and exchange phases matches the prediction $q^{\star} \!=\! -1$, and the exchange phase is consistent with $\varphi_{\rm exc} \!=\! \pi$, see Fig.~\ref{fig:ChernInsulator:ABPhase:N70}.
We conclude that sufficiently large lattice systems allow for faithful extraction of the exchange phase.
However, the system sizes needed are on the order of a few hundred lattice sites.

\section{Concluding discussions}

In this work, we introduce an interferometric protocol designed to directly probe the geometric phase of anyonic excitations in two-dimensional topological systems. Our approach makes use of impurities possessing two internal states, which interact with the surrounding medium. Specifically, in a fractional quantum Hall liquid, these impurities become bound to quasihole excitations. By precisely manipulating the impurities -- such as with optical tweezers -- one can implement impurity-based interferometry to resolve the braiding phase characteristic of anyonic statistics. The proposed experimental scheme is compatible with current state-of-the-art platforms, including ultracold atomic systems and potentially van der Waals heterostructures. Furthermore, our protocol is extensible to non-Abelian anyons, enabling full reconstruction of the associated Wilson loop matrix.

We performed numerical investigations to determine the system sizes necessary for reliably extracting the geometric phase via our interferometric protocol. Using a non-interacting Chern insulator as a benchmark, we found that systems comprising at least several hundred lattice sites are required to suppress detrimental edge effects -- that is, the system dimensions must significantly exceed the magnetic length $\ell_B$. Given that the relevant length scales are comparable in strongly interacting fractional Chern insulators, we infer that similarly large systems are essential for accurately measuring the anyonic exchange phase. However, simulating interacting lattice systems of this scale remains a formidable challenge, particularly for bosonic models. Consequently, our numerical studies of the interferometric scheme within the fractional quantum Hall regime of the Hofstadter-Hubbard model~\cite{Soerensen2005,Hafezi2007,Palmer2008,Moeller2009,Huegel2017,He2017,Dong2018,Raciunas2018,Repellin2020,Macaluso2020,Wang2022,Palm2022,Palm2024} were constrained to small system sizes and yielded inconclusive results at this stage, due to pronounced edge effects. As quantum simulators continue to scale up, they offer a promising avenue for exploring regimes that remain out of reach for classical numerical methods.

Our work paves the way toward anyon braiding experiments in solids or quantum gas microscopes with direct spatiotemporal control, also aiming for more intricate braiding processes involving non-Abelian anyons.
In particular, we expect similar protocols to be useful in future investigations of the quasiholes of Moore and Read's Pfaffian state~\cite{Moore1991}.
Ultimately, progress towards non-Abelian anyon braiding is speculated to give access to topologically protected quantum computation~\cite{Kitaev2003}.
While this is currently still out of reach for any existing platform, braiding Abelian Laughlin quasiholes in quantum simulators will be an important step towards this ambitious goal.

\vspace{0.5cm}
\begin{acknowledgments}
	The authors thank M.~Aidelsburger, E.~Demler, M.~Greiner, A.~\.Imamo\u{g}lu, A.~Nielsen, L.~Vanderstraeten, and B.~Wang for fruitful discussions.
    F.A.P., N.M., and F.G. acknowledge funding by the Deutsche Forschungsgemeinschaft (DFG, German Research Foundation) under Germany's Excellence Strategy – EXC-2111 – 390814868, and via Research Unit FOR 2414 under project number 277974659, and from the European Research Council (ERC) under the European Union’s Horizon 2020 research and innovation programm (Grant Agreement no 948141) – ERC Starting Grant SimUcQuam.
    Work in Brussels was financially supported by the FRS-FNRS (Belgium), the ERC Grant LATIS, the EOS project CHEQS, and the Fondation ULB.
\end{acknowledgments}

%
%  Bib
%
% \bibliographystyle[longbibliography]{apsrev4-2}
%\bibliography{literature}

\begin{thebibliography}{48}%
	\makeatletter
	\providecommand \@ifxundefined [1]{%
		\@ifx{#1\undefined}
	}%
	\providecommand \@ifnum [1]{%
		\ifnum #1\expandafter \@firstoftwo
		\else \expandafter \@secondoftwo
		\fi
	}%
	\providecommand \@ifx [1]{%
		\ifx #1\expandafter \@firstoftwo
		\else \expandafter \@secondoftwo
		\fi
	}%
	\providecommand \natexlab [1]{#1}%
	\providecommand \enquote  [1]{``#1''}%
	\providecommand \bibnamefont  [1]{#1}%
	\providecommand \bibfnamefont [1]{#1}%
	\providecommand \citenamefont [1]{#1}%
	\providecommand \href@noop [0]{\@secondoftwo}%
	\providecommand \href [0]{\begingroup \@sanitize@url \@href}%
	\providecommand \@href[1]{\@@startlink{#1}\@@href}%
	\providecommand \@@href[1]{\endgroup#1\@@endlink}%
	\providecommand \@sanitize@url [0]{\catcode `\\12\catcode `\$12\catcode
		`\&12\catcode `\#12\catcode `\^12\catcode `\_12\catcode `\%12\relax}%
	\providecommand \@@startlink[1]{}%
	\providecommand \@@endlink[0]{}%
	\providecommand \url  [0]{\begingroup\@sanitize@url \@url }%
	\providecommand \@url [1]{\endgroup\@href {#1}{\urlprefix }}%
	\providecommand \urlprefix  [0]{URL }%
	\providecommand \Eprint [0]{\href }%
	\providecommand \doibase [0]{https://doi.org/}%
	\providecommand \selectlanguage [0]{\@gobble}%
	\providecommand \bibinfo  [0]{\@secondoftwo}%
	\providecommand \bibfield  [0]{\@secondoftwo}%
	\providecommand \translation [1]{[#1]}%
	\providecommand \BibitemOpen [0]{}%
	\providecommand \bibitemStop [0]{}%
	\providecommand \bibitemNoStop [0]{.\EOS\space}%
	\providecommand \EOS [0]{\spacefactor3000\relax}%
	\providecommand \BibitemShut  [1]{\csname bibitem#1\endcsname}%
	\let\auto@bib@innerbib\@empty
	%</preamble>
	\bibitem [{\citenamefont {Nayak}\ \emph {et~al.}(2008)\citenamefont {Nayak},
		\citenamefont {Simon}, \citenamefont {Stern}, \citenamefont {Freedman},\ and\
		\citenamefont {Sarma}}]{Nayak2008}%
	\BibitemOpen
	\bibfield  {author} {\bibinfo {author} {\bibfnamefont {C.}~\bibnamefont
			{Nayak}}, \bibinfo {author} {\bibfnamefont {S.~H.}\ \bibnamefont {Simon}},
		\bibinfo {author} {\bibfnamefont {A.}~\bibnamefont {Stern}}, \bibinfo
		{author} {\bibfnamefont {M.}~\bibnamefont {Freedman}},\ and\ \bibinfo
		{author} {\bibfnamefont {S.~D.}\ \bibnamefont {Sarma}},\ }\bibfield  {title}
	{\bibinfo {title} {{Non-Abelian anyons and topological quantum
				computation}},\ }\href {https://doi.org/10.1103/revmodphys.80.1083}
	{\bibfield  {journal} {\bibinfo  {journal} {Reviews of Modern Physics}\
		}\textbf {\bibinfo {volume} {80}},\ \bibinfo {pages} {1083} (\bibinfo {year}
		{2008})}\BibitemShut {NoStop}%
	\bibitem [{\citenamefont {Kobayashi}\ and\ \citenamefont
		{Hashisaka}(2021)}]{kobayashi2021shot}%
	\BibitemOpen
	\bibfield  {author} {\bibinfo {author} {\bibfnamefont {K.}~\bibnamefont
			{Kobayashi}}\ and\ \bibinfo {author} {\bibfnamefont {M.}~\bibnamefont
			{Hashisaka}},\ }\bibfield  {title} {\bibinfo {title} {{Shot Noise in
				Mesoscopic Systems: From Single Particles to Quantum Liquids}},\ }\href
	{https://doi.org/10.7566/jpsj.90.102001} {\bibfield  {journal} {\bibinfo
			{journal} {Journal of the Physical Society of Japan}\ }\textbf {\bibinfo
			{volume} {90}},\ \bibinfo {pages} {102001} (\bibinfo {year}
		{2021})}\BibitemShut {NoStop}%
	\bibitem [{\citenamefont {Jonckheere}\ \emph {et~al.}(2023)\citenamefont
		{Jonckheere}, \citenamefont {Rech}, \citenamefont {Gr{\'{e}}maud},\ and\
		\citenamefont {Martin}}]{jonckheere2023anyonic}%
	\BibitemOpen
	\bibfield  {author} {\bibinfo {author} {\bibfnamefont {T.}~\bibnamefont
			{Jonckheere}}, \bibinfo {author} {\bibfnamefont {J.}~\bibnamefont {Rech}},
		\bibinfo {author} {\bibfnamefont {B.}~\bibnamefont {Gr{\'{e}}maud}},\ and\
		\bibinfo {author} {\bibfnamefont {T.}~\bibnamefont {Martin}},\ }\bibfield
	{title} {\bibinfo {title} {{Anyonic Statistics Revealed by the Hong-Ou-Mandel
				Dip for Fractional Excitations}},\ }\href
	{https://doi.org/10.1103/physrevlett.130.186203} {\bibfield  {journal}
		{\bibinfo  {journal} {Physical Review Letters}\ }\textbf {\bibinfo {volume}
			{130}},\ \bibinfo {pages} {186203} (\bibinfo {year} {2023})}\BibitemShut
	{NoStop}%
	\bibitem [{\citenamefont {Ronen}\ \emph {et~al.}(2021)\citenamefont {Ronen},
		\citenamefont {Werkmeister}, \citenamefont {Najafabadi}, \citenamefont
		{Pierce}, \citenamefont {Anderson}, \citenamefont {Shin}, \citenamefont
		{Lee}, \citenamefont {Lee}, \citenamefont {Johnson}, \citenamefont
		{Watanabe}, \citenamefont {Taniguchi}, \citenamefont {Yacoby},\ and\
		\citenamefont {Kim}}]{ronen2021aharonov}%
	\BibitemOpen
	\bibfield  {author} {\bibinfo {author} {\bibfnamefont {Y.}~\bibnamefont
			{Ronen}}, \bibinfo {author} {\bibfnamefont {T.}~\bibnamefont {Werkmeister}},
		\bibinfo {author} {\bibfnamefont {D.~H.}\ \bibnamefont {Najafabadi}},
		\bibinfo {author} {\bibfnamefont {A.~T.}\ \bibnamefont {Pierce}}, \bibinfo
		{author} {\bibfnamefont {L.~E.}\ \bibnamefont {Anderson}}, \bibinfo {author}
		{\bibfnamefont {Y.~J.}\ \bibnamefont {Shin}}, \bibinfo {author}
		{\bibfnamefont {S.~Y.}\ \bibnamefont {Lee}}, \bibinfo {author} {\bibfnamefont
			{Y.~H.}\ \bibnamefont {Lee}}, \bibinfo {author} {\bibfnamefont
			{B.}~\bibnamefont {Johnson}}, \bibinfo {author} {\bibfnamefont
			{K.}~\bibnamefont {Watanabe}}, \bibinfo {author} {\bibfnamefont
			{T.}~\bibnamefont {Taniguchi}}, \bibinfo {author} {\bibfnamefont
			{A.}~\bibnamefont {Yacoby}},\ and\ \bibinfo {author} {\bibfnamefont
			{P.}~\bibnamefont {Kim}},\ }\bibfield  {title} {\bibinfo {title}
		{{Aharonov{\textendash}Bohm effect in graphene-based
				Fabry{\textendash}P{\'{e}}rot quantum Hall interferometers}},\ }\href
	{https://doi.org/10.1038/s41565-021-00861-z} {\bibfield  {journal} {\bibinfo
			{journal} {Nature Nanotechnology}\ }\textbf {\bibinfo {volume} {16}},\
		\bibinfo {pages} {563} (\bibinfo {year} {2021})}\BibitemShut {NoStop}%
	\bibitem [{\citenamefont {Ruelle}\ \emph {et~al.}(2023)\citenamefont {Ruelle},
		\citenamefont {Frigerio}, \citenamefont {Berroir}, \citenamefont
		{Pla{\c{c}}ais}, \citenamefont {Rech}, \citenamefont {Cavanna}, \citenamefont
		{Gennser}, \citenamefont {Jin},\ and\ \citenamefont
		{F{\`{e}}ve}}]{ruelle2023comparing}%
	\BibitemOpen
	\bibfield  {author} {\bibinfo {author} {\bibfnamefont {M.}~\bibnamefont
			{Ruelle}}, \bibinfo {author} {\bibfnamefont {E.}~\bibnamefont {Frigerio}},
		\bibinfo {author} {\bibfnamefont {J.-M.}\ \bibnamefont {Berroir}}, \bibinfo
		{author} {\bibfnamefont {B.}~\bibnamefont {Pla{\c{c}}ais}}, \bibinfo {author}
		{\bibfnamefont {J.}~\bibnamefont {Rech}}, \bibinfo {author} {\bibfnamefont
			{A.}~\bibnamefont {Cavanna}}, \bibinfo {author} {\bibfnamefont
			{U.}~\bibnamefont {Gennser}}, \bibinfo {author} {\bibfnamefont
			{Y.}~\bibnamefont {Jin}},\ and\ \bibinfo {author} {\bibfnamefont
			{G.}~\bibnamefont {F{\`{e}}ve}},\ }\bibfield  {title} {\bibinfo {title}
		{{Comparing Fractional Quantum Hall Laughlin and Jain Topological Orders with
				the Anyon Collider}},\ }\href {https://doi.org/10.1103/physrevx.13.011031}
	{\bibfield  {journal} {\bibinfo  {journal} {Physical Review X}\ }\textbf
		{\bibinfo {volume} {13}},\ \bibinfo {pages} {011031} (\bibinfo {year}
		{2023})}\BibitemShut {NoStop}%
	\bibitem [{\citenamefont {Glidic}\ \emph {et~al.}(2023)\citenamefont {Glidic},
		\citenamefont {Maillet}, \citenamefont {Aassime}, \citenamefont {Piquard},
		\citenamefont {Cavanna}, \citenamefont {Gennser}, \citenamefont {Jin},
		\citenamefont {Anthore},\ and\ \citenamefont {Pierre}}]{glidic2023cross}%
	\BibitemOpen
	\bibfield  {author} {\bibinfo {author} {\bibfnamefont {P.}~\bibnamefont
			{Glidic}}, \bibinfo {author} {\bibfnamefont {O.}~\bibnamefont {Maillet}},
		\bibinfo {author} {\bibfnamefont {A.}~\bibnamefont {Aassime}}, \bibinfo
		{author} {\bibfnamefont {C.}~\bibnamefont {Piquard}}, \bibinfo {author}
		{\bibfnamefont {A.}~\bibnamefont {Cavanna}}, \bibinfo {author} {\bibfnamefont
			{U.}~\bibnamefont {Gennser}}, \bibinfo {author} {\bibfnamefont
			{Y.}~\bibnamefont {Jin}}, \bibinfo {author} {\bibfnamefont {A.}~\bibnamefont
			{Anthore}},\ and\ \bibinfo {author} {\bibfnamefont {F.}~\bibnamefont
			{Pierre}},\ }\bibfield  {title} {\bibinfo {title} {{Cross-Correlation
				Investigation of Anyon Statistics in the $\nu=1/3$ and $\nu=2/5$ Fractional
				Quantum Hall States}},\ }\href {https://doi.org/10.1103/physrevx.13.011030}
	{\bibfield  {journal} {\bibinfo  {journal} {Physical Review X}\ }\textbf
		{\bibinfo {volume} {13}},\ \bibinfo {pages} {011030} (\bibinfo {year}
		{2023})}\BibitemShut {NoStop}%
	\bibitem [{\citenamefont {Feldman}\ and\ \citenamefont
		{Halperin}(2021)}]{feldman2021fractional}%
	\BibitemOpen
	\bibfield  {author} {\bibinfo {author} {\bibfnamefont {D.~E.}\ \bibnamefont
			{Feldman}}\ and\ \bibinfo {author} {\bibfnamefont {B.~I.}\ \bibnamefont
			{Halperin}},\ }\bibfield  {title} {\bibinfo {title} {{Fractional charge and
				fractional statistics in the quantum Hall effects}},\ }\href
	{https://doi.org/10.1088/1361-6633/ac03aa} {\bibfield  {journal} {\bibinfo
			{journal} {Reports on Progress in Physics}\ }\textbf {\bibinfo {volume}
			{84}},\ \bibinfo {pages} {076501} (\bibinfo {year} {2021})}\BibitemShut
	{NoStop}%
	\bibitem [{\citenamefont {Lin}\ \emph {et~al.}(2021)\citenamefont {Lin},
		\citenamefont {Hashisaka}, \citenamefont {Akiho}, \citenamefont {Muraki},\
		and\ \citenamefont {Fujisawa}}]{lin2021quantized}%
	\BibitemOpen
	\bibfield  {author} {\bibinfo {author} {\bibfnamefont {C.}~\bibnamefont
			{Lin}}, \bibinfo {author} {\bibfnamefont {M.}~\bibnamefont {Hashisaka}},
		\bibinfo {author} {\bibfnamefont {T.}~\bibnamefont {Akiho}}, \bibinfo
		{author} {\bibfnamefont {K.}~\bibnamefont {Muraki}},\ and\ \bibinfo {author}
		{\bibfnamefont {T.}~\bibnamefont {Fujisawa}},\ }\bibfield  {title} {\bibinfo
		{title} {{Quantized charge fractionalization at quantum Hall Y junctions in
				the disorder dominated regime}},\ }\bibfield  {journal} {\bibinfo  {journal}
		{Nature Communications}\ }\textbf {\bibinfo {volume} {12}},\ \href
	{https://doi.org/10.1038/s41467-020-20395-7} {10.1038/s41467-020-20395-7}
	(\bibinfo {year} {2021})\BibitemShut {NoStop}%
	\bibitem [{\citenamefont {Bartolomei}\ \emph {et~al.}(2020)\citenamefont
		{Bartolomei}, \citenamefont {Kumar}, \citenamefont {Bisognin}, \citenamefont
		{Marguerite}, \citenamefont {Berroir}, \citenamefont {Bocquillon},
		\citenamefont {Pla{\c{c}}ais}, \citenamefont {Cavanna}, \citenamefont {Dong},
		\citenamefont {Gennser}, \citenamefont {Jin},\ and\ \citenamefont
		{F{\`{e}}ve}}]{bartolomei2020fractional}%
	\BibitemOpen
	\bibfield  {author} {\bibinfo {author} {\bibfnamefont {H.}~\bibnamefont
			{Bartolomei}}, \bibinfo {author} {\bibfnamefont {M.}~\bibnamefont {Kumar}},
		\bibinfo {author} {\bibfnamefont {R.}~\bibnamefont {Bisognin}}, \bibinfo
		{author} {\bibfnamefont {A.}~\bibnamefont {Marguerite}}, \bibinfo {author}
		{\bibfnamefont {J.-M.}\ \bibnamefont {Berroir}}, \bibinfo {author}
		{\bibfnamefont {E.}~\bibnamefont {Bocquillon}}, \bibinfo {author}
		{\bibfnamefont {B.}~\bibnamefont {Pla{\c{c}}ais}}, \bibinfo {author}
		{\bibfnamefont {A.}~\bibnamefont {Cavanna}}, \bibinfo {author} {\bibfnamefont
			{Q.}~\bibnamefont {Dong}}, \bibinfo {author} {\bibfnamefont {U.}~\bibnamefont
			{Gennser}}, \bibinfo {author} {\bibfnamefont {Y.}~\bibnamefont {Jin}},\ and\
		\bibinfo {author} {\bibfnamefont {G.}~\bibnamefont {F{\`{e}}ve}},\ }\bibfield
	{title} {\bibinfo {title} {Fractional statistics in anyon collisions},\
	}\href {https://doi.org/10.1126/science.aaz5601} {\bibfield  {journal}
		{\bibinfo  {journal} {Science}\ }\textbf {\bibinfo {volume} {368}},\ \bibinfo
		{pages} {173} (\bibinfo {year} {2020})}\BibitemShut {NoStop}%
	\bibitem [{\citenamefont {Nakamura}\ \emph {et~al.}(2020)\citenamefont
		{Nakamura}, \citenamefont {Liang}, \citenamefont {Gardner},\ and\
		\citenamefont {Manfra}}]{nakamura2020direct}%
	\BibitemOpen
	\bibfield  {author} {\bibinfo {author} {\bibfnamefont {J.}~\bibnamefont
			{Nakamura}}, \bibinfo {author} {\bibfnamefont {S.}~\bibnamefont {Liang}},
		\bibinfo {author} {\bibfnamefont {G.~C.}\ \bibnamefont {Gardner}},\ and\
		\bibinfo {author} {\bibfnamefont {M.~J.}\ \bibnamefont {Manfra}},\ }\bibfield
	{title} {\bibinfo {title} {Direct observation of anyonic braiding
			statistics},\ }\href {https://doi.org/10.1038/s41567-020-1019-1} {\bibfield
		{journal} {\bibinfo  {journal} {Nature Physics}\ }\textbf {\bibinfo {volume}
			{16}},\ \bibinfo {pages} {931} (\bibinfo {year} {2020})}\BibitemShut
	{NoStop}%
	\bibitem [{\citenamefont {Lee}\ \emph {et~al.}(2023)\citenamefont {Lee},
		\citenamefont {Hong}, \citenamefont {Alkalay}, \citenamefont {Schiller},
		\citenamefont {Umansky}, \citenamefont {Heiblum}, \citenamefont {Oreg},\ and\
		\citenamefont {Sim}}]{lee2023partitioning}%
	\BibitemOpen
	\bibfield  {author} {\bibinfo {author} {\bibfnamefont {J.-Y.~M.}\
			\bibnamefont {Lee}}, \bibinfo {author} {\bibfnamefont {C.}~\bibnamefont
			{Hong}}, \bibinfo {author} {\bibfnamefont {T.}~\bibnamefont {Alkalay}},
		\bibinfo {author} {\bibfnamefont {N.}~\bibnamefont {Schiller}}, \bibinfo
		{author} {\bibfnamefont {V.}~\bibnamefont {Umansky}}, \bibinfo {author}
		{\bibfnamefont {M.}~\bibnamefont {Heiblum}}, \bibinfo {author} {\bibfnamefont
			{Y.}~\bibnamefont {Oreg}},\ and\ \bibinfo {author} {\bibfnamefont {H.-S.}\
			\bibnamefont {Sim}},\ }\bibfield  {title} {\bibinfo {title} {{Partitioning of
				diluted anyons reveals their braiding statistics}},\ }\href
	{https://doi.org/10.1038/s41586-023-05883-2} {\bibfield  {journal} {\bibinfo
			{journal} {Nature}\ }\textbf {\bibinfo {volume} {617}},\ \bibinfo {pages}
		{277} (\bibinfo {year} {2023})}\BibitemShut {NoStop}%
	\bibitem [{\citenamefont {Satzinger}\ \emph {et~al.}(2021)\citenamefont
		{Satzinger}, \citenamefont {Liu}, \citenamefont {Smith}, \citenamefont
		{Knapp}, \citenamefont {Newman}, \citenamefont {Jones}, \citenamefont {Chen},
		\citenamefont {Quintana}, \citenamefont {Mi}, \citenamefont {Dunsworth} \emph
		{et~al.}}]{satzinger2021realizing}%
	\BibitemOpen
	\bibfield  {author} {\bibinfo {author} {\bibfnamefont {K.~J.}\ \bibnamefont
			{Satzinger}}, \bibinfo {author} {\bibfnamefont {Y.-J.}\ \bibnamefont {Liu}},
		\bibinfo {author} {\bibfnamefont {A.}~\bibnamefont {Smith}}, \bibinfo
		{author} {\bibfnamefont {C.}~\bibnamefont {Knapp}}, \bibinfo {author}
		{\bibfnamefont {M.}~\bibnamefont {Newman}}, \bibinfo {author} {\bibfnamefont
			{C.}~\bibnamefont {Jones}}, \bibinfo {author} {\bibfnamefont
			{Z.}~\bibnamefont {Chen}}, \bibinfo {author} {\bibfnamefont {C.}~\bibnamefont
			{Quintana}}, \bibinfo {author} {\bibfnamefont {X.}~\bibnamefont {Mi}},
		\bibinfo {author} {\bibfnamefont {A.}~\bibnamefont {Dunsworth}}, \emph
		{et~al.},\ }\bibfield  {title} {\bibinfo {title} {Realizing topologically
			ordered states on a quantum processor},\ }\href
	{https://doi.org/10.1126/science.abi8378} {\bibfield  {journal} {\bibinfo
			{journal} {Science}\ }\textbf {\bibinfo {volume} {374}},\ \bibinfo {pages}
		{1237} (\bibinfo {year} {2021})}\BibitemShut {NoStop}%
	\bibitem [{\citenamefont {Iqbal}\ \emph {et~al.}(2024)\citenamefont {Iqbal},
		\citenamefont {Tantivasadakarn}, \citenamefont {Verresen}, \citenamefont
		{Campbell}, \citenamefont {Dreiling}, \citenamefont {Figgatt}, \citenamefont
		{Gaebler}, \citenamefont {Johansen}, \citenamefont {Mills}, \citenamefont
		{Moses}, \citenamefont {Pino}, \citenamefont {Ransford}, \citenamefont
		{Rowe}, \citenamefont {Siegfried}, \citenamefont {Stutz}, \citenamefont
		{Foss-Feig}, \citenamefont {Vishwanath},\ and\ \citenamefont
		{Dreyer}}]{Iqbal2024}%
	\BibitemOpen
	\bibfield  {author} {\bibinfo {author} {\bibfnamefont {M.}~\bibnamefont
			{Iqbal}}, \bibinfo {author} {\bibfnamefont {N.}~\bibnamefont
			{Tantivasadakarn}}, \bibinfo {author} {\bibfnamefont {R.}~\bibnamefont
			{Verresen}}, \bibinfo {author} {\bibfnamefont {S.~L.}\ \bibnamefont
			{Campbell}}, \bibinfo {author} {\bibfnamefont {J.~M.}\ \bibnamefont
			{Dreiling}}, \bibinfo {author} {\bibfnamefont {C.}~\bibnamefont {Figgatt}},
		\bibinfo {author} {\bibfnamefont {J.~P.}\ \bibnamefont {Gaebler}}, \bibinfo
		{author} {\bibfnamefont {J.}~\bibnamefont {Johansen}}, \bibinfo {author}
		{\bibfnamefont {M.}~\bibnamefont {Mills}}, \bibinfo {author} {\bibfnamefont
			{S.~A.}\ \bibnamefont {Moses}}, \bibinfo {author} {\bibfnamefont {J.~M.}\
			\bibnamefont {Pino}}, \bibinfo {author} {\bibfnamefont {A.}~\bibnamefont
			{Ransford}}, \bibinfo {author} {\bibfnamefont {M.}~\bibnamefont {Rowe}},
		\bibinfo {author} {\bibfnamefont {P.}~\bibnamefont {Siegfried}}, \bibinfo
		{author} {\bibfnamefont {R.~P.}\ \bibnamefont {Stutz}}, \bibinfo {author}
		{\bibfnamefont {M.}~\bibnamefont {Foss-Feig}}, \bibinfo {author}
		{\bibfnamefont {A.}~\bibnamefont {Vishwanath}},\ and\ \bibinfo {author}
		{\bibfnamefont {H.}~\bibnamefont {Dreyer}},\ }\bibfield  {title} {\bibinfo
		{title} {{Non-Abelian topological order and anyons on a trapped-ion
				processor}},\ }\href {https://doi.org/10.1038/s41586-023-06934-4} {\bibfield
		{journal} {\bibinfo  {journal} {Nature}\ }\textbf {\bibinfo {volume} {626}},\
		\bibinfo {pages} {505} (\bibinfo {year} {2024})}\BibitemShut {NoStop}%
	\bibitem [{\citenamefont {Andersen}\ \emph {et~al.}(2023)\citenamefont
		{Andersen}, \citenamefont {Lensky}, \citenamefont {Kechedzhi}, \citenamefont
		{Drozdov}, \citenamefont {Bengtsson}, \citenamefont {Hong}, \citenamefont
		{Morvan}, \citenamefont {Mi}, \citenamefont {Opremcak}, \citenamefont
		{Acharya} \emph {et~al.}}]{andersen2023observation}%
	\BibitemOpen
	\bibfield  {author} {\bibinfo {author} {\bibfnamefont {T.~I.}\ \bibnamefont
			{Andersen}}, \bibinfo {author} {\bibfnamefont {Y.~D.}\ \bibnamefont
			{Lensky}}, \bibinfo {author} {\bibfnamefont {K.}~\bibnamefont {Kechedzhi}},
		\bibinfo {author} {\bibfnamefont {I.~K.}\ \bibnamefont {Drozdov}}, \bibinfo
		{author} {\bibfnamefont {A.}~\bibnamefont {Bengtsson}}, \bibinfo {author}
		{\bibfnamefont {S.}~\bibnamefont {Hong}}, \bibinfo {author} {\bibfnamefont
			{A.}~\bibnamefont {Morvan}}, \bibinfo {author} {\bibfnamefont
			{X.}~\bibnamefont {Mi}}, \bibinfo {author} {\bibfnamefont {A.}~\bibnamefont
			{Opremcak}}, \bibinfo {author} {\bibfnamefont {R.}~\bibnamefont {Acharya}},
		\emph {et~al.},\ }\bibfield  {title} {\bibinfo {title} {{Non-Abelian braiding
				of graph vertices in a superconducting processor}},\ }\href
	{https://doi.org/10.1038/s41586-023-05954-4} {\bibfield  {journal} {\bibinfo
			{journal} {Nature}\ }\textbf {\bibinfo {volume} {618}},\ \bibinfo {pages}
		{264} (\bibinfo {year} {2023})}\BibitemShut {NoStop}%
	\bibitem [{\citenamefont {Cooper}\ \emph {et~al.}(2019)\citenamefont {Cooper},
		\citenamefont {Dalibard},\ and\ \citenamefont
		{Spielman}}]{cooper2019topological}%
	\BibitemOpen
	\bibfield  {author} {\bibinfo {author} {\bibfnamefont {N.~R.}\ \bibnamefont
			{Cooper}}, \bibinfo {author} {\bibfnamefont {J.}~\bibnamefont {Dalibard}},\
		and\ \bibinfo {author} {\bibfnamefont {I.~B.}\ \bibnamefont {Spielman}},\
	}\bibfield  {title} {\bibinfo {title} {Topological bands for ultracold
			atoms},\ }\href {https://doi.org/10.1103/revmodphys.91.015005} {\bibfield
		{journal} {\bibinfo  {journal} {Reviews of Modern Physics}\ }\textbf
		{\bibinfo {volume} {91}},\ \bibinfo {pages} {015005} (\bibinfo {year}
		{2019})}\BibitemShut {NoStop}%
	\bibitem [{\citenamefont {Ozawa}\ \emph {et~al.}(2019)\citenamefont {Ozawa},
		\citenamefont {Price}, \citenamefont {Amo}, \citenamefont {Goldman},
		\citenamefont {Hafezi}, \citenamefont {Lu}, \citenamefont {Rechtsman},
		\citenamefont {Schuster}, \citenamefont {Simon}, \citenamefont {Zilberberg},\
		and\ \citenamefont {Carusotto}}]{Ozawa2019}%
	\BibitemOpen
	\bibfield  {author} {\bibinfo {author} {\bibfnamefont {T.}~\bibnamefont
			{Ozawa}}, \bibinfo {author} {\bibfnamefont {H.~M.}\ \bibnamefont {Price}},
		\bibinfo {author} {\bibfnamefont {A.}~\bibnamefont {Amo}}, \bibinfo {author}
		{\bibfnamefont {N.}~\bibnamefont {Goldman}}, \bibinfo {author} {\bibfnamefont
			{M.}~\bibnamefont {Hafezi}}, \bibinfo {author} {\bibfnamefont
			{L.}~\bibnamefont {Lu}}, \bibinfo {author} {\bibfnamefont {M.~C.}\
			\bibnamefont {Rechtsman}}, \bibinfo {author} {\bibfnamefont {D.}~\bibnamefont
			{Schuster}}, \bibinfo {author} {\bibfnamefont {J.}~\bibnamefont {Simon}},
		\bibinfo {author} {\bibfnamefont {O.}~\bibnamefont {Zilberberg}},\ and\
		\bibinfo {author} {\bibfnamefont {I.}~\bibnamefont {Carusotto}},\ }\bibfield
	{title} {\bibinfo {title} {Topological photonics},\ }\href
	{https://doi.org/10.1103/revmodphys.91.015006} {\bibfield  {journal}
		{\bibinfo  {journal} {Reviews of Modern Physics}\ }\textbf {\bibinfo {volume}
			{91}},\ \bibinfo {pages} {015006} (\bibinfo {year} {2019})}\BibitemShut
	{NoStop}%
	\bibitem [{\citenamefont {Tai}\ \emph {et~al.}(2017)\citenamefont {Tai},
		\citenamefont {Lukin}, \citenamefont {Rispoli}, \citenamefont {Schittko},
		\citenamefont {Menke}, \citenamefont {Borgnia}, \citenamefont {Preiss},
		\citenamefont {Grusdt}, \citenamefont {Kaufman},\ and\ \citenamefont
		{Greiner}}]{tai2017microscopy}%
	\BibitemOpen
	\bibfield  {author} {\bibinfo {author} {\bibfnamefont {M.~E.}\ \bibnamefont
			{Tai}}, \bibinfo {author} {\bibfnamefont {A.}~\bibnamefont {Lukin}}, \bibinfo
		{author} {\bibfnamefont {M.}~\bibnamefont {Rispoli}}, \bibinfo {author}
		{\bibfnamefont {R.}~\bibnamefont {Schittko}}, \bibinfo {author}
		{\bibfnamefont {T.}~\bibnamefont {Menke}}, \bibinfo {author} {\bibfnamefont
			{D.}~\bibnamefont {Borgnia}}, \bibinfo {author} {\bibfnamefont {P.~M.}\
			\bibnamefont {Preiss}}, \bibinfo {author} {\bibfnamefont {F.}~\bibnamefont
			{Grusdt}}, \bibinfo {author} {\bibfnamefont {A.~M.}\ \bibnamefont
			{Kaufman}},\ and\ \bibinfo {author} {\bibfnamefont {M.}~\bibnamefont
			{Greiner}},\ }\bibfield  {title} {\bibinfo {title} {{Microscopy of the
				interacting Harper{\textendash}Hofstadter model in the two-body limit}},\
	}\href {https://doi.org/10.1038/nature22811} {\bibfield  {journal} {\bibinfo
			{journal} {Nature}\ }\textbf {\bibinfo {volume} {546}},\ \bibinfo {pages}
		{519} (\bibinfo {year} {2017})}\BibitemShut {NoStop}%
	\bibitem [{\citenamefont {L{\'{e}}onard}\ \emph {et~al.}(2023)\citenamefont
		{L{\'{e}}onard}, \citenamefont {Kim}, \citenamefont {Kwan}, \citenamefont
		{Segura}, \citenamefont {Grusdt}, \citenamefont {Repellin}, \citenamefont
		{Goldman},\ and\ \citenamefont {Greiner}}]{Leonard2023}%
	\BibitemOpen
	\bibfield  {author} {\bibinfo {author} {\bibfnamefont {J.}~\bibnamefont
			{L{\'{e}}onard}}, \bibinfo {author} {\bibfnamefont {S.}~\bibnamefont {Kim}},
		\bibinfo {author} {\bibfnamefont {J.}~\bibnamefont {Kwan}}, \bibinfo {author}
		{\bibfnamefont {P.}~\bibnamefont {Segura}}, \bibinfo {author} {\bibfnamefont
			{F.}~\bibnamefont {Grusdt}}, \bibinfo {author} {\bibfnamefont
			{C.}~\bibnamefont {Repellin}}, \bibinfo {author} {\bibfnamefont
			{N.}~\bibnamefont {Goldman}},\ and\ \bibinfo {author} {\bibfnamefont
			{M.}~\bibnamefont {Greiner}},\ }\bibfield  {title} {\bibinfo {title}
		{{Realization of a fractional quantum Hall state with ultracold atoms}},\
	}\href {https://doi.org/10.1038/s41586-023-06122-4} {\bibfield  {journal}
		{\bibinfo  {journal} {Nature}\ }\textbf {\bibinfo {volume} {619}},\ \bibinfo
		{pages} {495} (\bibinfo {year} {2023})}\BibitemShut {NoStop}%
	\bibitem [{\citenamefont {Clark}\ \emph {et~al.}(2020)\citenamefont {Clark},
		\citenamefont {Schine}, \citenamefont {Baum}, \citenamefont {Jia},\ and\
		\citenamefont {Simon}}]{Clark2020}%
	\BibitemOpen
	\bibfield  {author} {\bibinfo {author} {\bibfnamefont {L.~W.}\ \bibnamefont
			{Clark}}, \bibinfo {author} {\bibfnamefont {N.}~\bibnamefont {Schine}},
		\bibinfo {author} {\bibfnamefont {C.}~\bibnamefont {Baum}}, \bibinfo {author}
		{\bibfnamefont {N.}~\bibnamefont {Jia}},\ and\ \bibinfo {author}
		{\bibfnamefont {J.}~\bibnamefont {Simon}},\ }\bibfield  {title} {\bibinfo
		{title} {{Observation of Laughlin states made of light}},\ }\href
	{https://doi.org/10.1038/s41586-020-2318-5} {\bibfield  {journal} {\bibinfo
			{journal} {Nature}\ }\textbf {\bibinfo {volume} {582}},\ \bibinfo {pages}
		{41} (\bibinfo {year} {2020})}\BibitemShut {NoStop}%
	\bibitem [{\citenamefont {Lunt}\ \emph {et~al.}(2024)\citenamefont {Lunt},
		\citenamefont {Hill}, \citenamefont {Reiter}, \citenamefont {Preiss},
		\citenamefont {Ga\l{}ka},\ and\ \citenamefont {Jochim}}]{Lunt2025}%
	\BibitemOpen
	\bibfield  {author} {\bibinfo {author} {\bibfnamefont {P.}~\bibnamefont
			{Lunt}}, \bibinfo {author} {\bibfnamefont {P.}~\bibnamefont {Hill}}, \bibinfo
		{author} {\bibfnamefont {J.}~\bibnamefont {Reiter}}, \bibinfo {author}
		{\bibfnamefont {P.~M.}\ \bibnamefont {Preiss}}, \bibinfo {author}
		{\bibfnamefont {M.}~\bibnamefont {Ga\l{}ka}},\ and\ \bibinfo {author}
		{\bibfnamefont {S.}~\bibnamefont {Jochim}},\ }\bibfield  {title} {\bibinfo
		{title} {Realization of a laughlin state of two rapidly rotating fermions},\
	}\href {https://doi.org/10.1103/PhysRevLett.133.253401} {\bibfield  {journal}
		{\bibinfo  {journal} {Physical Review Letters}\ }\textbf {\bibinfo {volume}
			{133}},\ \bibinfo {pages} {253401} (\bibinfo {year} {2024})}\BibitemShut
	{NoStop}%
	\bibitem [{\citenamefont {Wang}\ \emph {et~al.}(2024)\citenamefont {Wang},
		\citenamefont {Liu}, \citenamefont {Chen}, \citenamefont {Chen},
		\citenamefont {Zhao}, \citenamefont {Ying}, \citenamefont {Shang},
		\citenamefont {Wang}, \citenamefont {Huo}, \citenamefont {Peng},
		\citenamefont {Zhu}, \citenamefont {Lu},\ and\ \citenamefont
		{Pan}}]{Wang2024}%
	\BibitemOpen
	\bibfield  {author} {\bibinfo {author} {\bibfnamefont {C.}~\bibnamefont
			{Wang}}, \bibinfo {author} {\bibfnamefont {F.-M.}\ \bibnamefont {Liu}},
		\bibinfo {author} {\bibfnamefont {M.-C.}\ \bibnamefont {Chen}}, \bibinfo
		{author} {\bibfnamefont {H.}~\bibnamefont {Chen}}, \bibinfo {author}
		{\bibfnamefont {X.-H.}\ \bibnamefont {Zhao}}, \bibinfo {author}
		{\bibfnamefont {C.}~\bibnamefont {Ying}}, \bibinfo {author} {\bibfnamefont
			{Z.-X.}\ \bibnamefont {Shang}}, \bibinfo {author} {\bibfnamefont {J.-W.}\
			\bibnamefont {Wang}}, \bibinfo {author} {\bibfnamefont {Y.-H.}\ \bibnamefont
			{Huo}}, \bibinfo {author} {\bibfnamefont {C.-Z.}\ \bibnamefont {Peng}},
		\bibinfo {author} {\bibfnamefont {X.}~\bibnamefont {Zhu}}, \bibinfo {author}
		{\bibfnamefont {C.-Y.}\ \bibnamefont {Lu}},\ and\ \bibinfo {author}
		{\bibfnamefont {J.-W.}\ \bibnamefont {Pan}},\ }\bibfield  {title} {\bibinfo
		{title} {{Realization of fractional quantum Hall state with interacting
				photons}},\ }\href {https://doi.org/10.1126/science.ado3912} {\bibfield
		{journal} {\bibinfo  {journal} {Science}\ }\textbf {\bibinfo {volume}
			{384}},\ \bibinfo {pages} {579} (\bibinfo {year} {2024})}\BibitemShut
	{NoStop}%
	\bibitem [{\citenamefont {Paredes}\ \emph {et~al.}(2001)\citenamefont
		{Paredes}, \citenamefont {Fedichev}, \citenamefont {Cirac},\ and\
		\citenamefont {Zoller}}]{paredes20011}%
	\BibitemOpen
	\bibfield  {author} {\bibinfo {author} {\bibfnamefont {B.}~\bibnamefont
			{Paredes}}, \bibinfo {author} {\bibfnamefont {P.}~\bibnamefont {Fedichev}},
		\bibinfo {author} {\bibfnamefont {J.~I.}\ \bibnamefont {Cirac}},\ and\
		\bibinfo {author} {\bibfnamefont {P.}~\bibnamefont {Zoller}},\ }\bibfield
	{title} {\bibinfo {title} {{$\frac{1}{2}$-Anyons in Small Atomic
				Bose-Einstein Condensates}},\ }\href
	{https://doi.org/10.1103/physrevlett.87.010402} {\bibfield  {journal}
		{\bibinfo  {journal} {Physical Review Letters}\ }\textbf {\bibinfo {volume}
			{87}},\ \bibinfo {pages} {010402} (\bibinfo {year} {2001})}\BibitemShut
	{NoStop}%
	\bibitem [{\citenamefont {Ra{\v{c}}i{\={u}}nas}\ \emph
		{et~al.}(2018)\citenamefont {Ra{\v{c}}i{\={u}}nas}, \citenamefont {Ünal},
		\citenamefont {Anisimovas},\ and\ \citenamefont {Eckardt}}]{Raciunas2018}%
	\BibitemOpen
	\bibfield  {author} {\bibinfo {author} {\bibfnamefont {M.}~\bibnamefont
			{Ra{\v{c}}i{\={u}}nas}}, \bibinfo {author} {\bibfnamefont {F.~N.}\
			\bibnamefont {Ünal}}, \bibinfo {author} {\bibfnamefont {E.}~\bibnamefont
			{Anisimovas}},\ and\ \bibinfo {author} {\bibfnamefont {A.}~\bibnamefont
			{Eckardt}},\ }\bibfield  {title} {\bibinfo {title} {{Creating, probing, and
				manipulating fractionally charged excitations of fractional Chern insulators
				in optical lattices}},\ }\href {https://doi.org/10.1103/physreva.98.063621}
	{\bibfield  {journal} {\bibinfo  {journal} {Physical Review A}\ }\textbf
		{\bibinfo {volume} {98}},\ \bibinfo {pages} {063621} (\bibinfo {year}
		{2018})}\BibitemShut {NoStop}%
	\bibitem [{\citenamefont {Macaluso}\ \emph {et~al.}(2020)\citenamefont
		{Macaluso}, \citenamefont {Comparin}, \citenamefont {Umucal{\i}lar},
		\citenamefont {Gerster}, \citenamefont {Montangero}, \citenamefont {Rizzi},\
		and\ \citenamefont {Carusotto}}]{Macaluso2020}%
	\BibitemOpen
	\bibfield  {author} {\bibinfo {author} {\bibfnamefont {E.}~\bibnamefont
			{Macaluso}}, \bibinfo {author} {\bibfnamefont {T.}~\bibnamefont {Comparin}},
		\bibinfo {author} {\bibfnamefont {R.~O.}\ \bibnamefont {Umucal{\i}lar}},
		\bibinfo {author} {\bibfnamefont {M.}~\bibnamefont {Gerster}}, \bibinfo
		{author} {\bibfnamefont {S.}~\bibnamefont {Montangero}}, \bibinfo {author}
		{\bibfnamefont {M.}~\bibnamefont {Rizzi}},\ and\ \bibinfo {author}
		{\bibfnamefont {I.}~\bibnamefont {Carusotto}},\ }\bibfield  {title} {\bibinfo
		{title} {{Charge and statistics of lattice quasiholes from density
				measurements: A tree tensor network study}},\ }\href
	{https://doi.org/10.1103/physrevresearch.2.013145} {\bibfield  {journal}
		{\bibinfo  {journal} {Physical Review Research}\ }\textbf {\bibinfo {volume}
			{2}},\ \bibinfo {pages} {013145} (\bibinfo {year} {2020})}\BibitemShut
	{NoStop}%
	\bibitem [{\citenamefont {Wang}\ \emph {et~al.}(2022)\citenamefont {Wang},
		\citenamefont {Dong},\ and\ \citenamefont {Eckardt}}]{Wang2022}%
	\BibitemOpen
	\bibfield  {author} {\bibinfo {author} {\bibfnamefont {B.}~\bibnamefont
			{Wang}}, \bibinfo {author} {\bibfnamefont {X.}~\bibnamefont {Dong}},\ and\
		\bibinfo {author} {\bibfnamefont {A.}~\bibnamefont {Eckardt}},\ }\bibfield
	{title} {\bibinfo {title} {{Measurable signatures of bosonic fractional Chern
				insulator states and their fractional excitations in a quantum-gas
				microscope}},\ }\href {https://doi.org/10.21468/scipostphys.12.3.095}
	{\bibfield  {journal} {\bibinfo  {journal} {{SciPost} Physics}\ }\textbf
		{\bibinfo {volume} {12}},\ \bibinfo {pages} {095} (\bibinfo {year}
		{2022})}\BibitemShut {NoStop}%
	\bibitem [{\citenamefont {Palm}\ \emph {et~al.}(2024)\citenamefont {Palm},
		\citenamefont {Kwan}, \citenamefont {Bakkali-Hassani}, \citenamefont
		{Greiner}, \citenamefont {Schollwöck}, \citenamefont {Goldman},\ and\
		\citenamefont {Grusdt}}]{Palm2024}%
	\BibitemOpen
	\bibfield  {author} {\bibinfo {author} {\bibfnamefont {F.~A.}\ \bibnamefont
			{Palm}}, \bibinfo {author} {\bibfnamefont {J.}~\bibnamefont {Kwan}}, \bibinfo
		{author} {\bibfnamefont {B.}~\bibnamefont {Bakkali-Hassani}}, \bibinfo
		{author} {\bibfnamefont {M.}~\bibnamefont {Greiner}}, \bibinfo {author}
		{\bibfnamefont {U.}~\bibnamefont {Schollwöck}}, \bibinfo {author}
		{\bibfnamefont {N.}~\bibnamefont {Goldman}},\ and\ \bibinfo {author}
		{\bibfnamefont {F.}~\bibnamefont {Grusdt}},\ }\bibfield  {title} {\bibinfo
		{title} {{Growing Extended Laughlin States in a Quantum Gas Microscope: A
				Patchwork Construction}},\ }\href
	{https://doi.org/10.1103/PhysRevResearch.6.013198} {\bibfield  {journal}
		{\bibinfo  {journal} {Physical Review Research}\ }\textbf {\bibinfo {volume}
			{6}},\ \bibinfo {pages} {013198} (\bibinfo {year} {2024})}\BibitemShut
	{NoStop}%
	\bibitem [{\citenamefont {Mostaan}\ \emph {et~al.}(2025)\citenamefont
		{Mostaan}, \citenamefont {Goldman}, \citenamefont {{\.I}mamo{\u{g}}lu},\ and\
		\citenamefont {Grusdt}}]{mostaan2025anyon}%
	\BibitemOpen
	\bibfield  {author} {\bibinfo {author} {\bibfnamefont {N.}~\bibnamefont
			{Mostaan}}, \bibinfo {author} {\bibfnamefont {N.}~\bibnamefont {Goldman}},
		\bibinfo {author} {\bibfnamefont {A.}~\bibnamefont {{\.I}mamo{\u{g}}lu}},\
		and\ \bibinfo {author} {\bibfnamefont {F.}~\bibnamefont {Grusdt}},\
	}\bibfield  {title} {\bibinfo {title} {Anyon-trions in atomically thin
			semiconductor heterostructures},\ }\href@noop {} {\bibfield  {journal}
		{\bibinfo  {journal} {arXiv preprint arXiv:2507.08933}\ } (\bibinfo {year}
		{2025})}\BibitemShut {NoStop}%
	\bibitem [{\citenamefont {Wagner}\ and\ \citenamefont
		{Neupert}(2025)}]{wagner2025sensing}%
	\BibitemOpen
	\bibfield  {author} {\bibinfo {author} {\bibfnamefont {G.}~\bibnamefont
			{Wagner}}\ and\ \bibinfo {author} {\bibfnamefont {T.}~\bibnamefont
			{Neupert}},\ }\bibfield  {title} {\bibinfo {title} {Sensing the binding and
			unbinding of anyons at impurities},\ }\href@noop {} {\bibfield  {journal}
		{\bibinfo  {journal} {arXiv preprint arXiv:2507.08928}\ } (\bibinfo {year}
		{2025})}\BibitemShut {NoStop}%
	\bibitem [{\citenamefont {Grusdt}\ \emph {et~al.}(2016)\citenamefont {Grusdt},
		\citenamefont {Yao}, \citenamefont {Abanin}, \citenamefont {Fleischhauer},\
		and\ \citenamefont {Demler}}]{Grusdt2016}%
	\BibitemOpen
	\bibfield  {author} {\bibinfo {author} {\bibfnamefont {F.}~\bibnamefont
			{Grusdt}}, \bibinfo {author} {\bibfnamefont {N.~Y.}\ \bibnamefont {Yao}},
		\bibinfo {author} {\bibfnamefont {D.}~\bibnamefont {Abanin}}, \bibinfo
		{author} {\bibfnamefont {M.}~\bibnamefont {Fleischhauer}},\ and\ \bibinfo
		{author} {\bibfnamefont {E.}~\bibnamefont {Demler}},\ }\bibfield  {title}
	{\bibinfo {title} {Interferometric measurements of many-body topological
			invariants using mobile impurities},\ }\href
	{https://doi.org/10.1038/ncomms11994} {\bibfield  {journal} {\bibinfo
			{journal} {Nature Communications}\ }\textbf {\bibinfo {volume} {7}},\
		\bibinfo {pages} {11994} (\bibinfo {year} {2016})}\BibitemShut {NoStop}%
	\bibitem [{sup()}]{supp}%
	\BibitemOpen
	\bibinfo {note} {See Supplemental Material for details on the impurity
		interferometry sequence and supporting data.}\BibitemShut {Stop}%
	\bibitem [{\citenamefont {Moore}\ and\ \citenamefont {Read}(1991)}]{Moore1991}%
	\BibitemOpen
	\bibfield  {author} {\bibinfo {author} {\bibfnamefont {G.}~\bibnamefont
			{Moore}}\ and\ \bibinfo {author} {\bibfnamefont {N.}~\bibnamefont {Read}},\
	}\bibfield  {title} {\bibinfo {title} {{Nonabelions in the fractional quantum
				hall effect}},\ }\href {https://doi.org/10.1016/0550-3213(91)90407-o}
	{\bibfield  {journal} {\bibinfo  {journal} {Nuclear Physics B}\ }\textbf
		{\bibinfo {volume} {360}},\ \bibinfo {pages} {362} (\bibinfo {year}
		{1991})}\BibitemShut {NoStop}%
	\bibitem [{\citenamefont {Barb{\'{e}}}\ \emph {et~al.}(2018)\citenamefont
		{Barb{\'{e}}}, \citenamefont {Ciamei}, \citenamefont {Pasquiou},
		\citenamefont {Reichsöllner}, \citenamefont {Schreck}, \citenamefont
		{{\.{Z}}uchowski},\ and\ \citenamefont {Hutson}}]{barbe2018observation}%
	\BibitemOpen
	\bibfield  {author} {\bibinfo {author} {\bibfnamefont {V.}~\bibnamefont
			{Barb{\'{e}}}}, \bibinfo {author} {\bibfnamefont {A.}~\bibnamefont {Ciamei}},
		\bibinfo {author} {\bibfnamefont {B.}~\bibnamefont {Pasquiou}}, \bibinfo
		{author} {\bibfnamefont {L.}~\bibnamefont {Reichsöllner}}, \bibinfo {author}
		{\bibfnamefont {F.}~\bibnamefont {Schreck}}, \bibinfo {author} {\bibfnamefont
			{P.~S.}\ \bibnamefont {{\.{Z}}uchowski}},\ and\ \bibinfo {author}
		{\bibfnamefont {J.~M.}\ \bibnamefont {Hutson}},\ }\bibfield  {title}
	{\bibinfo {title} {{Observation of Feshbach resonances between alkali and
				closed-shell atoms}},\ }\href {https://doi.org/10.1038/s41567-018-0169-x}
	{\bibfield  {journal} {\bibinfo  {journal} {Nature Physics}\ }\textbf
		{\bibinfo {volume} {14}},\ \bibinfo {pages} {881} (\bibinfo {year}
		{2018})}\BibitemShut {NoStop}%
	\bibitem [{\citenamefont {Heinz}\ \emph {et~al.}(2020)\citenamefont {Heinz},
		\citenamefont {Park}, \citenamefont {{\v{S}}anti{\'{c}}}, \citenamefont
		{Trautmann}, \citenamefont {Porsev}, \citenamefont {Safronova}, \citenamefont
		{Bloch},\ and\ \citenamefont {Blatt}}]{heinz2020state}%
	\BibitemOpen
	\bibfield  {author} {\bibinfo {author} {\bibfnamefont {A.}~\bibnamefont
			{Heinz}}, \bibinfo {author} {\bibfnamefont {A.}~\bibnamefont {Park}},
		\bibinfo {author} {\bibfnamefont {N.}~\bibnamefont {{\v{S}}anti{\'{c}}}},
		\bibinfo {author} {\bibfnamefont {J.}~\bibnamefont {Trautmann}}, \bibinfo
		{author} {\bibfnamefont {S.}~\bibnamefont {Porsev}}, \bibinfo {author}
		{\bibfnamefont {M.}~\bibnamefont {Safronova}}, \bibinfo {author}
		{\bibfnamefont {I.}~\bibnamefont {Bloch}},\ and\ \bibinfo {author}
		{\bibfnamefont {S.}~\bibnamefont {Blatt}},\ }\bibfield  {title} {\bibinfo
		{title} {{State-Dependent Optical Lattices for the Strontium Optical
				Qubit}},\ }\href {https://doi.org/10.1103/physrevlett.124.203201} {\bibfield
		{journal} {\bibinfo  {journal} {Physical Review Letters}\ }\textbf {\bibinfo
			{volume} {124}},\ \bibinfo {pages} {203201} (\bibinfo {year}
		{2020})}\BibitemShut {NoStop}%
	\bibitem [{\citenamefont {Popert}\ \emph {et~al.}(2022)\citenamefont {Popert},
		\citenamefont {Shimazaki}, \citenamefont {Kroner}, \citenamefont {Watanabe},
		\citenamefont {Taniguchi}, \citenamefont {Imamoglu},\ and\ \citenamefont
		{Smolenski}}]{popert2022optical}%
	\BibitemOpen
	\bibfield  {author} {\bibinfo {author} {\bibfnamefont {A.}~\bibnamefont
			{Popert}}, \bibinfo {author} {\bibfnamefont {Y.}~\bibnamefont {Shimazaki}},
		\bibinfo {author} {\bibfnamefont {M.}~\bibnamefont {Kroner}}, \bibinfo
		{author} {\bibfnamefont {K.}~\bibnamefont {Watanabe}}, \bibinfo {author}
		{\bibfnamefont {T.}~\bibnamefont {Taniguchi}}, \bibinfo {author}
		{\bibfnamefont {A.}~\bibnamefont {Imamoglu}},\ and\ \bibinfo {author}
		{\bibfnamefont {T.}~\bibnamefont {Smolenski}},\ }\bibfield  {title} {\bibinfo
		{title} {Optical sensing of fractional quantum hall effect in graphene},\
	}\href@noop {} {\bibfield  {journal} {\bibinfo  {journal} {Nano Letters}\
		}\textbf {\bibinfo {volume} {22}},\ \bibinfo {pages} {7363} (\bibinfo {year}
		{2022})}\BibitemShut {NoStop}%
	\bibitem [{\citenamefont {Cui}\ \emph {et~al.}(2024)\citenamefont {Cui},
		\citenamefont {Hu}, \citenamefont {Zhao}, \citenamefont {Ma}, \citenamefont
		{Jin}, \citenamefont {Zhang}, \citenamefont {Watanabe}, \citenamefont
		{Taniguchi}, \citenamefont {Shan}, \citenamefont {Mak} \emph
		{et~al.}}]{cui2024interlayer}%
	\BibitemOpen
	\bibfield  {author} {\bibinfo {author} {\bibfnamefont {H.}~\bibnamefont
			{Cui}}, \bibinfo {author} {\bibfnamefont {Q.}~\bibnamefont {Hu}}, \bibinfo
		{author} {\bibfnamefont {X.}~\bibnamefont {Zhao}}, \bibinfo {author}
		{\bibfnamefont {L.}~\bibnamefont {Ma}}, \bibinfo {author} {\bibfnamefont
			{F.}~\bibnamefont {Jin}}, \bibinfo {author} {\bibfnamefont {Q.}~\bibnamefont
			{Zhang}}, \bibinfo {author} {\bibfnamefont {K.}~\bibnamefont {Watanabe}},
		\bibinfo {author} {\bibfnamefont {T.}~\bibnamefont {Taniguchi}}, \bibinfo
		{author} {\bibfnamefont {J.}~\bibnamefont {Shan}}, \bibinfo {author}
		{\bibfnamefont {K.~F.}\ \bibnamefont {Mak}}, \emph {et~al.},\ }\bibfield
	{title} {\bibinfo {title} {Interlayer fermi polarons of excited exciton
			states in quantizing magnetic fields},\ }\href@noop {} {\bibfield  {journal}
		{\bibinfo  {journal} {Nano Letters}\ }\textbf {\bibinfo {volume} {24}},\
		\bibinfo {pages} {7077} (\bibinfo {year} {2024})}\BibitemShut {NoStop}%
	\bibitem [{\citenamefont {Gao}\ \emph {et~al.}(2025)\citenamefont {Gao},
		\citenamefont {Ghafariasl}, \citenamefont {Mehrabad}, \citenamefont {Huang},
		\citenamefont {Zhang}, \citenamefont {Session}, \citenamefont {Upadhyay},
		\citenamefont {Ma}, \citenamefont {Alshalan}, \citenamefont {Forero} \emph
		{et~al.}}]{gao2025probing}%
	\BibitemOpen
	\bibfield  {author} {\bibinfo {author} {\bibfnamefont {B.}~\bibnamefont
			{Gao}}, \bibinfo {author} {\bibfnamefont {M.}~\bibnamefont {Ghafariasl}},
		\bibinfo {author} {\bibfnamefont {M.~J.}\ \bibnamefont {Mehrabad}}, \bibinfo
		{author} {\bibfnamefont {T.-S.}\ \bibnamefont {Huang}}, \bibinfo {author}
		{\bibfnamefont {L.}~\bibnamefont {Zhang}}, \bibinfo {author} {\bibfnamefont
			{D.}~\bibnamefont {Session}}, \bibinfo {author} {\bibfnamefont
			{P.}~\bibnamefont {Upadhyay}}, \bibinfo {author} {\bibfnamefont
			{R.}~\bibnamefont {Ma}}, \bibinfo {author} {\bibfnamefont {G.}~\bibnamefont
			{Alshalan}}, \bibinfo {author} {\bibfnamefont {D.~G.~S.}\ \bibnamefont
			{Forero}}, \emph {et~al.},\ }\bibfield  {title} {\bibinfo {title} {Probing
			quantum anomalous hall states in twisted bilayer wse2 via attractive polaron
			spectroscopy},\ }\href@noop {} {\bibfield  {journal} {\bibinfo  {journal}
			{arXiv preprint arXiv:2504.11530}\ } (\bibinfo {year} {2025})}\BibitemShut
	{NoStop}%
	\bibitem [{\citenamefont {Kapit}\ \emph {et~al.}(2012)\citenamefont {Kapit},
		\citenamefont {Ginsparg},\ and\ \citenamefont {Mueller}}]{Kapit2012}%
	\BibitemOpen
	\bibfield  {author} {\bibinfo {author} {\bibfnamefont {E.}~\bibnamefont
			{Kapit}}, \bibinfo {author} {\bibfnamefont {P.}~\bibnamefont {Ginsparg}},\
		and\ \bibinfo {author} {\bibfnamefont {E.}~\bibnamefont {Mueller}},\
	}\bibfield  {title} {\bibinfo {title} {{Non-Abelian Braiding of Lattice
				Bosons}},\ }\href {https://doi.org/10.1103/physrevlett.108.066802} {\bibfield
		{journal} {\bibinfo  {journal} {Physical Review Letters}\ }\textbf {\bibinfo
			{volume} {108}},\ \bibinfo {pages} {066802} (\bibinfo {year}
		{2012})}\BibitemShut {NoStop}%
	\bibitem [{\citenamefont {Resta}(1994)}]{Resta1994}%
	\BibitemOpen
	\bibfield  {author} {\bibinfo {author} {\bibfnamefont {R.}~\bibnamefont
			{Resta}},\ }\bibfield  {title} {\bibinfo {title} {{Macroscopic polarization
				in crystalline dielectrics: the geometric phase approach}},\ }\href
	{https://doi.org/10.1103/revmodphys.66.899} {\bibfield  {journal} {\bibinfo
			{journal} {Reviews of Modern Physics}\ }\textbf {\bibinfo {volume} {66}},\
		\bibinfo {pages} {899} (\bibinfo {year} {1994})}\BibitemShut {NoStop}%
	\bibitem [{\citenamefont {S{\o}rensen}\ \emph {et~al.}(2005)\citenamefont
		{S{\o}rensen}, \citenamefont {Demler},\ and\ \citenamefont
		{Lukin}}]{Soerensen2005}%
	\BibitemOpen
	\bibfield  {author} {\bibinfo {author} {\bibfnamefont {A.~S.}\ \bibnamefont
			{S{\o}rensen}}, \bibinfo {author} {\bibfnamefont {E.}~\bibnamefont
			{Demler}},\ and\ \bibinfo {author} {\bibfnamefont {M.~D.}\ \bibnamefont
			{Lukin}},\ }\bibfield  {title} {\bibinfo {title} {{Fractional Quantum Hall
				States of Atoms in Optical Lattices}},\ }\href
	{https://doi.org/10.1103/physrevlett.94.086803} {\bibfield  {journal}
		{\bibinfo  {journal} {Physical Review Letters}\ }\textbf {\bibinfo {volume}
			{94}},\ \bibinfo {pages} {086803} (\bibinfo {year} {2005})}\BibitemShut
	{NoStop}%
	\bibitem [{\citenamefont {Hafezi}\ \emph {et~al.}(2007)\citenamefont {Hafezi},
		\citenamefont {S{\o}rensen}, \citenamefont {Demler},\ and\ \citenamefont
		{Lukin}}]{Hafezi2007}%
	\BibitemOpen
	\bibfield  {author} {\bibinfo {author} {\bibfnamefont {M.}~\bibnamefont
			{Hafezi}}, \bibinfo {author} {\bibfnamefont {A.~S.}\ \bibnamefont
			{S{\o}rensen}}, \bibinfo {author} {\bibfnamefont {E.}~\bibnamefont
			{Demler}},\ and\ \bibinfo {author} {\bibfnamefont {M.~D.}\ \bibnamefont
			{Lukin}},\ }\bibfield  {title} {\bibinfo {title} {{Fractional quantum Hall
				effect in optical lattices}},\ }\href
	{https://doi.org/10.1103/physreva.76.023613} {\bibfield  {journal} {\bibinfo
			{journal} {Physical Review A}\ }\textbf {\bibinfo {volume} {76}},\ \bibinfo
		{pages} {023613} (\bibinfo {year} {2007})}\BibitemShut {NoStop}%
	\bibitem [{\citenamefont {Palmer}\ \emph {et~al.}(2008)\citenamefont {Palmer},
		\citenamefont {Klein},\ and\ \citenamefont {Jaksch}}]{Palmer2008}%
	\BibitemOpen
	\bibfield  {author} {\bibinfo {author} {\bibfnamefont {R.~N.}\ \bibnamefont
			{Palmer}}, \bibinfo {author} {\bibfnamefont {A.}~\bibnamefont {Klein}},\ and\
		\bibinfo {author} {\bibfnamefont {D.}~\bibnamefont {Jaksch}},\ }\bibfield
	{title} {\bibinfo {title} {{Optical lattice quantum Hall effect}},\ }\href
	{https://doi.org/10.1103/physreva.78.013609} {\bibfield  {journal} {\bibinfo
			{journal} {Physical Review A}\ }\textbf {\bibinfo {volume} {78}},\ \bibinfo
		{pages} {013609} (\bibinfo {year} {2008})}\BibitemShut {NoStop}%
	\bibitem [{\citenamefont {Möller}\ and\ \citenamefont
		{Cooper}(2009)}]{Moeller2009}%
	\BibitemOpen
	\bibfield  {author} {\bibinfo {author} {\bibfnamefont {G.}~\bibnamefont
			{Möller}}\ and\ \bibinfo {author} {\bibfnamefont {N.~R.}\ \bibnamefont
			{Cooper}},\ }\bibfield  {title} {\bibinfo {title} {{Composite Fermion Theory
				for Bosonic Quantum Hall States on Lattices}},\ }\href
	{https://doi.org/10.1103/physrevlett.103.105303} {\bibfield  {journal}
		{\bibinfo  {journal} {Physical Review Letters}\ }\textbf {\bibinfo {volume}
			{103}},\ \bibinfo {pages} {105303} (\bibinfo {year} {2009})}\BibitemShut
	{NoStop}%
	\bibitem [{\citenamefont {Hügel}\ \emph {et~al.}(2017)\citenamefont {Hügel},
		\citenamefont {Strand}, \citenamefont {Werner},\ and\ \citenamefont
		{Pollet}}]{Huegel2017}%
	\BibitemOpen
	\bibfield  {author} {\bibinfo {author} {\bibfnamefont {D.}~\bibnamefont
			{Hügel}}, \bibinfo {author} {\bibfnamefont {H.~U.~R.}\ \bibnamefont
			{Strand}}, \bibinfo {author} {\bibfnamefont {P.}~\bibnamefont {Werner}},\
		and\ \bibinfo {author} {\bibfnamefont {L.}~\bibnamefont {Pollet}},\
	}\bibfield  {title} {\bibinfo {title} {{Anisotropic Harper-Hofstadter-Mott
				model: Competition between condensation and magnetic fields}},\ }\href
	{https://doi.org/10.1103/physrevb.96.054431} {\bibfield  {journal} {\bibinfo
			{journal} {Physical Review B}\ }\textbf {\bibinfo {volume} {96}},\ \bibinfo
		{pages} {054431} (\bibinfo {year} {2017})}\BibitemShut {NoStop}%
	\bibitem [{\citenamefont {He}\ \emph {et~al.}(2017)\citenamefont {He},
		\citenamefont {Grusdt}, \citenamefont {Kaufman}, \citenamefont {Greiner},\
		and\ \citenamefont {Vishwanath}}]{He2017}%
	\BibitemOpen
	\bibfield  {author} {\bibinfo {author} {\bibfnamefont {Y.-C.}\ \bibnamefont
			{He}}, \bibinfo {author} {\bibfnamefont {F.}~\bibnamefont {Grusdt}}, \bibinfo
		{author} {\bibfnamefont {A.}~\bibnamefont {Kaufman}}, \bibinfo {author}
		{\bibfnamefont {M.}~\bibnamefont {Greiner}},\ and\ \bibinfo {author}
		{\bibfnamefont {A.}~\bibnamefont {Vishwanath}},\ }\bibfield  {title}
	{\bibinfo {title} {{Realizing and adiabatically preparing bosonic integer and
				fractional quantum Hall states in optical lattices}},\ }\href
	{https://doi.org/10.1103/physrevb.96.201103} {\bibfield  {journal} {\bibinfo
			{journal} {Physical Review B}\ }\textbf {\bibinfo {volume} {96}},\ \bibinfo
		{pages} {201103(R)} (\bibinfo {year} {2017})}\BibitemShut {NoStop}%
	\bibitem [{\citenamefont {Dong}\ \emph {et~al.}(2018)\citenamefont {Dong},
		\citenamefont {Grushin}, \citenamefont {Motruk},\ and\ \citenamefont
		{Pollmann}}]{Dong2018}%
	\BibitemOpen
	\bibfield  {author} {\bibinfo {author} {\bibfnamefont {X.-Y.}\ \bibnamefont
			{Dong}}, \bibinfo {author} {\bibfnamefont {A.~G.}\ \bibnamefont {Grushin}},
		\bibinfo {author} {\bibfnamefont {J.}~\bibnamefont {Motruk}},\ and\ \bibinfo
		{author} {\bibfnamefont {F.}~\bibnamefont {Pollmann}},\ }\bibfield  {title}
	{\bibinfo {title} {{Charge Excitation Dynamics in Bosonic Fractional Chern
				Insulators}},\ }\href {https://doi.org/10.1103/physrevlett.121.086401}
	{\bibfield  {journal} {\bibinfo  {journal} {Physical Review Letters}\
		}\textbf {\bibinfo {volume} {121}},\ \bibinfo {pages} {086401} (\bibinfo
		{year} {2018})}\BibitemShut {NoStop}%
	\bibitem [{\citenamefont {Repellin}\ \emph {et~al.}(2020)\citenamefont
		{Repellin}, \citenamefont {L{\'{e}}onard},\ and\ \citenamefont
		{Goldman}}]{Repellin2020}%
	\BibitemOpen
	\bibfield  {author} {\bibinfo {author} {\bibfnamefont {C.}~\bibnamefont
			{Repellin}}, \bibinfo {author} {\bibfnamefont {J.}~\bibnamefont
			{L{\'{e}}onard}},\ and\ \bibinfo {author} {\bibfnamefont {N.}~\bibnamefont
			{Goldman}},\ }\bibfield  {title} {\bibinfo {title} {{Fractional Chern
				insulators of few bosons in a box: Hall plateaus from center-of-mass drifts
				and density profiles}},\ }\href {https://doi.org/10.1103/physreva.102.063316}
	{\bibfield  {journal} {\bibinfo  {journal} {Physical Review A}\ }\textbf
		{\bibinfo {volume} {102}},\ \bibinfo {pages} {063316} (\bibinfo {year}
		{2020})}\BibitemShut {NoStop}%
	\bibitem [{\citenamefont {Palm}\ \emph {et~al.}(2022)\citenamefont {Palm},
		\citenamefont {Mardazad}, \citenamefont {Bohrdt}, \citenamefont
		{Schollw\"ock},\ and\ \citenamefont {Grusdt}}]{Palm2022}%
	\BibitemOpen
	\bibfield  {author} {\bibinfo {author} {\bibfnamefont {F.~A.}\ \bibnamefont
			{Palm}}, \bibinfo {author} {\bibfnamefont {S.}~\bibnamefont {Mardazad}},
		\bibinfo {author} {\bibfnamefont {A.}~\bibnamefont {Bohrdt}}, \bibinfo
		{author} {\bibfnamefont {U.}~\bibnamefont {Schollw\"ock}},\ and\ \bibinfo
		{author} {\bibfnamefont {F.}~\bibnamefont {Grusdt}},\ }\bibfield  {title}
	{\bibinfo {title} {{Snapshot-based detection of $\nu=\frac{1}{2}$ Laughlin
				states: Coupled chains and central charge}},\ }\href
	{https://doi.org/10.1103/PhysRevB.106.L081108} {\bibfield  {journal}
		{\bibinfo  {journal} {Physical Review B}\ }\textbf {\bibinfo {volume}
			{106}},\ \bibinfo {pages} {L081108} (\bibinfo {year} {2022})}\BibitemShut
	{NoStop}%
	\bibitem [{\citenamefont {Kitaev}(2003)}]{Kitaev2003}%
	\BibitemOpen
	\bibfield  {author} {\bibinfo {author} {\bibfnamefont {A.~Y.}\ \bibnamefont
			{Kitaev}},\ }\bibfield  {title} {\bibinfo {title} {{Fault-tolerant quantum
				computation by anyons}},\ }\href
	{https://doi.org/10.1016/s0003-4916(02)00018-0} {\bibfield  {journal}
		{\bibinfo  {journal} {Annals of Physics}\ }\textbf {\bibinfo {volume}
			{303}},\ \bibinfo {pages} {2} (\bibinfo {year} {2003})}\BibitemShut {NoStop}%
\end{thebibliography}


%%%%%%%%%%%%%%%%%%%%%%%%%%%%%%%%%%%%%%%%%%%%%%%%%%%%%%%%%%%%%%%%%%%%%%%%%%%%%%%%%%%%%%%%%%%%%%%%%%%
%%%%%%%%%%%%%%%%%%%%%%%%%%%%%%%%%%%%%%%%%%%%%%%%%%%%%%%%%%%%%%%%%%%%%%%%%%%%%%%%%%%%%%%%%%%%%%%%%%%

\renewcommand{\theequation}{S\arabic{equation}}
\renewcommand{\thefigure}{S\arabic{figure}}
\renewcommand{\thetable}{S\arabic{table}}

\onecolumngrid

%\newpage

\setcounter{equation}{0}
\setcounter{figure}{0}
\setcounter{table}{0}
\setcounter{section}{0}

%%%%%%%%%%%%%%%%%%%%%%%%%%%%%%%%%%%%%%%%%%%%%%%%%%%%
\clearpage

\section*{SUPPLEMENTAL MATERIAL:\\ Interferometric Braiding of Anyons in Chern Insulators}
\setcounter{page}{1}
\begin{center}
F.~A.~Palm,$^{1,2,3}$ 
N.~Mostaan,$^{1,2,3}$
N. Goldman,$^{3,4,5}$ and
F. Grusdt,$^{1,2}$ \\
\textit{$^{1}$ Department of Physics and Arnold Sommerfeld Center for Theoretical Physics (ASC), Ludwig-Maximilians-Universit\"at M\"unchen, Theresienstr. 37, D-80333 M\"unchen, Germany}\\
\textit{$^{2}$ Munich Center for Quantum Science and Technology (MCQST), Schellingstr. 4, D-80799 M\"unchen, Germany}\\
\textit{$^{3}$ CENOLI, Universit\'e Libre de Bruxelles, CP 231, Campus Plaine, B-1050 Brussels, Belgium}\\
\textit{$^{4}$ Laboratoire Kastler Brossel, Coll\`ege de France, CNRS, ENS-Universit\'e PSL, Sorbonne Universit\'e, 11 Place Marcelin Berthelot, 75005 Paris, France}\\
\textit{$^{5}$ International Solvay Institutes, 1050 Brussels, Belgium}
\end{center}

\section{Interferometric Sequence}
\subsection{One impurity}
Here we elaborate on the evolution of the full quantum state through the measurement protocol.
The system is initially prepared in the quantum state $\ket{\Psi_{\rm init}} = \ket{\vc{r} \uparrow}\ \ket{\mathrm{qh}}_{\mathrm{QH}}$ as in Eq.~\eqref{eq:psi_i:1ptcl} in the main text.
Application of a $\pi/2-$pulse brings the state to $\ket{\vc{r} +} \ket{\mt{qh}}_{\mt{QH}}$, with the following form in terms of $\ket{\uparrow},\,\ket{\downarrow}$-states,
\begin{equation}\label{supp:eq:psi_1:1ptcl}
    \ket{\Psi_1} = \frac{1}{\sqrt{2}} \ket{\vc{r} \uparrow} \ket{\mt{qh}}_{\mt{QH}} + \frac{1}{\sqrt{2}} \ket{\vc{r} \downarrow} \ket{\mt{qh}}_{\mt{QH}}.
\end{equation}
The action of translating the $\ket{\uparrow}$ impurity from $\vc{r}$ to $\vc{r}^{\prime}$ on the quasihole state is represented by the unitary operator $\hat{U}_{\vc{r}\to\vc{r}^{\prime}}$.
The resulting quantum state takes the form
\begin{equation}\label{supp:eq:psi_2:1ptcl}
    \ket{\Psi_2} = \frac{1}{\sqrt{2}} \ket{\vc{r}^{\prime}\uparrow} 
    \hat{U}_{\vc{r}\to\vc{r}^{\prime}}\ket{\mt{qh}}_{\mt{QH}} + \frac{1}{\sqrt{2}} \ket{\vc{r} \downarrow} \ket{\mt{qh}}_{\mt{QH}}.
\end{equation}
In the next step, a $\pi-$pulse flips the impurity spin, leading to the state
\begin{equation}\label{supp:eq:psi_2:1ptcl}
    \ket{\Psi_3} = \frac{1}{\sqrt{2}} \ket{\vc{r}^{\prime}\downarrow} 
    \hat{U}_{\vc{r}\to\vc{r}^{\prime}}\ket{\mt{qh}}_{\mt{QH}} + \frac{1}{\sqrt{2}} \ket{\vc{r} \uparrow} \ket{\mt{qh}}_{\mt{QH}}.
\end{equation}
The $\ket{\uparrow}$-impurity at $\vc{r}$ is transferred to $\vc{r}^{\prime}$ over the complementary part of the closed path in Fig.~\ref{fig:SketchABPhase}(a), resulting in a unitary transformation $\hat{U}^{'}_{\vc{r}\to\vc{r}^{\prime}}$ on the quasihole state. The total state after this operation reads
\begin{equation}\label{supp:eq:psi_2:1ptcl}
    \ket{\Psi_4} = \frac{1}{\sqrt{2}} \ket{\vc{r}^{\prime}\downarrow} 
    \hat{U}_{\vc{r}\to\vc{r}^{\prime}}\ket{\mt{qh}}_{\mt{QH}} + \frac{1}{\sqrt{2}} \ket{\vc{r}^{\prime} \uparrow} \hat{U}^{'}_{\vc{r}\to\vc{r}^{\prime}} \ket{\mt{qh}}_{\mt{QH}}.
\end{equation}
Finally, application of a $\pi/2-$pulse with a control phase $\phi_{c}$ to $\ket{\Psi_4}$ implements the transformations $\ket{\uparrow}\to \ket{+_{\phi_c}}=1/\sqrt{2} \big( \ket{\uparrow} + e^{i\phi_c} \ket{\downarrow} \big)$ and $\ket{\downarrow}\to \ket{-_{\phi_c}}= 1/\sqrt{2} \big( e^{-i\phi_c} \ket{\uparrow} - \ket{\downarrow} \big)$, resulting in the state $\ket{\Psi_{\mathrm{final}}}$ given by
\begin{equation}
    \ket{\Psi_{\rm final}} = \frac{1}{\sqrt{2}} \ket{\vc{r}^{\prime} -_{\phi_c}} \,\hat{U}_{\vc{r}\to\vc{r}^{\prime}}\ket{\mt{qh}}_{\mt{QH}} + \frac{1}{\sqrt{2}} \ket{\vc{r}^{\prime}+ _{\phi_c}} \hat{U}^{'}_{\vc{r}\to\vc{r}^{\prime}}\ket{\mt{qh}}_{\mt{QH}}.
\end{equation}
The unnormalized post-measurement state of the quantum Hall system $\ket{\Psi_{\mt{pm}}}=\big(\bra{\vc{r}^{\prime}\uparrow}\otimes\mathbb{I}_{\mt{QH}}\big)\ket{\Psi_{\rm final}}$ is 
\begin{equation}
    \ket{\Psi_{\mt{pm}}} = \frac{1}{2} e^{-i\phi_c} \, \hat{U}_{\vc{r}\to\vc{r}^{\prime}}\ket{\mt{qh}}_{\mt{QH}} + \frac{1}{2} \hat{U}^{'}_{\vc{r}\to\vc{r}^{\prime}}\ket{\mt{qh}}_{\mt{QH}}
    = \frac{1}{2} \hat{U}^{'}_{\vc{r}\to\vc{r}^{\prime}} \Big( \mathbb{I} + e^{-i\phi_c} \, \hat{U}^{'\dagger}_{\vc{r}\to\vc{r}^{\prime}} \hat{U}_{\vc{r}\to\vc{r}^{\prime}} \Big) \ket{\mt{qh}}_{\mt{QH}}.
\end{equation}
The unitary operator $\hat{U}^{'\dagger}_{\vc{r}\to\vc{r}^{\prime}} \hat{U}_{\vc{r}\to\vc{r}^{\prime}} $ in the second term in parantheses acts as a phase factor that a quasihole picks up upon traversing the closed path $\vc{r}\to\vc{r}^{\prime}\to\vc{r}$.
Assuming that this path preserves the spatial symmetries of the system, the dynamical contribution to the phase cancels and we are left with the geometric phase $\mt{exp}[i\varphi_{\mt{geo}}]$.
The norm of $\ket{\Psi_{\mt{pm}}}$ gives the measurement probability $p_{\uparrow}=\left|1/2 \left(1+\mt{exp}[i(\varphi_{\mt{geo}}-\phi_c)] \right)\right|^2 = \mt{cos}^2\left((\varphi_{\mt{geo}}-\phi_c)/2\right)$ for finding the impurity in the $\ket{\uparrow}$ state.

\subsection{Two impurities}
Here we elaborate on the evolution of the full quantum state through the measurement protocol.
The system is initially prepared in the quantum state $\ket{\Psi_i} = \ket{\vc{r}_1 \uparrow,\vc{r}_2\uparrow}\ \ket{\mathrm{2-qh}}_{\mathrm{QH}}$ as in Eq.~\eqref{eq:psi_i} in the main text.
Application of a $\pi/2-$pulse brings the state to $\ket{\vc{r}_1 +,\vc{r}_2 +} \ket{2-\mt{qh}}_{\mt{QH}}$, with the following form in terms of $\ket{\uparrow},\,\ket{\downarrow}$-states,
\begin{equation}\label{supp:eq:psi_1}
\begin{split}
    \ket{\Psi_1} = & \frac{1}{2} \, \ket{\vc{r}_1 \uparrow,\vc{r}_2 \uparrow} \ket{2-\mt{qh}}_{\mt{QH}} + \frac{1}{2} \, \ket{\vc{r}_1 \uparrow,\vc{r}_2 \downarrow} \ket{2-\mt{qh}}_{\mt{QH}} \\
    & + \frac{1}{2} \, \ket{\vc{r}_1 \downarrow,\vc{r}_2 \uparrow} \ket{2-\mt{qh}}_{\mt{QH}} + \frac{1}{2} \, \ket{\vc{r}_1 \downarrow,\vc{r}_2 \downarrow} \ket{2-\mt{qh}}_{\mt{QH}} \, .
\end{split}
\end{equation}
The operation of translating a single quasihole from position $\vc{r}$ to $\vc{r}'$ along a given path $\mathcal{C}$, while keeping all other quasiholes fixed at their initial positions, is represented by the operator $\hat{U}^{(\mathcal{C})}_{\vc{r}\to\vc{r}'}$, defined as
\begin{equation}\label{eq:UDef}
    \hat{U}^{(\mathcal{C})}_{\vc{r}\to\vc{r}'} 
    \equiv 
    e^{-i e^{*}/\hbar c \int_{\mathcal{C}} d\vc{x} \cdot \vc{\mathcal{A}}(\vc{x}) } \,
    \hat{T}_{\vc{r}\to\vc{r}'} \, ,
\end{equation}
where $\hat{T}_{\vc{r}\to\vc{r}'}$ denotes the translation operator, and the exponential prefactor arises from the coupling of the quasihole to the gauge field $\vc{\mathcal{A}}(\vc{x})$. Although $\hat{U}^{(\mathcal{C})}_{\vc{r}\to\vc{r}'}$ in Eq.~\eqref{eq:UDef} is defined in continuous space, its generalization to lattice systems is straightforward. The gauge field coupling to the quasihole consists of two contributions,
$\vc{\mathcal{A}}(\vc{x}) = \vc{\mathcal{A}}_{\mathrm{bg}}(\vc{x}) + \vc{\mathcal{A}}_{\mathrm{exc}}(\vc{x})$,
where $\vc{\mathcal{A}}_{\mathrm{bg}}(\vc{x})$ and $\vc{\mathcal{A}}_{\mathrm{exc}}(\vc{x})$ correspond to the background magnetic field and to the exchange flux of the existing quasiholes, respectively. Consequently, the geometric phase $\varphi_{\mathrm{geo}}$ acquired during the exchange process contains two distinct contributions, $\varphi_{\mathrm{AB}}$ and $\varphi_{\mathrm{exc}}$, arising from $\vc{\mathcal{A}}_{\mathrm{bg}}(\vc{x})$ and $\vc{\mathcal{A}}_{\mathrm{exc}}(\vc{x})$, respectively (see Eq.~\eqref{eq:geometricPhase} in the main text). Hereafter, since the exchange path $\mathcal{C}$ is specified in the main text, we omit the superscript $(\mathcal{C})$ in the definition of $\hat{U}$.

Given the definition of $\hat{U}_{\vc{r}\to\vc{r}'}$ above, the action of translating two $\ket{\uparrow}$-impurities from $(\vc{r}_1,\vc{r}_2)$ to $(\vc{r}_3,\vc{r}_4)$ on the quasihole states is represented by the unitary operator $\hat{U}_{\vc{r}_1\to\vc{r}_3}\hat{U}_{\vc{r}_2\to\vc{r}_4}$. The resulting quantum state takes the form
\begin{equation}\label{supp:eq:psi_2}
\begin{split}
    \ket{\Psi_2} = & \frac{1}{2} \, \ket{\vc{r}_3 \uparrow,\vc{r}_4 \uparrow} 
    \hat{U}_{\vc{r}_1\to\vc{r}_3}\hat{U}_{\vc{r}_2\to\vc{r}_4}\ket{2-\mt{qh}}_{\mt{QH}} + \frac{1}{2} \, \ket{\vc{r}_3 \uparrow,\vc{r}_2 \downarrow} 
    \hat{U}_{\vc{r}_1\to\vc{r}_3}\ket{2-\mt{qh}}_{\mt{QH}} \\ & + \frac{1}{2} \, \ket{\vc{r}_1 \downarrow,\vc{r}_4 \uparrow} 
    \hat{U}_{\vc{r}_2\to\vc{r}_4}\ket{2-\mt{qh}}_{\mt{QH}} + \frac{1}{2} \, \ket{\vc{r}_1 \downarrow,\vc{r}_2 \downarrow} 
    \ket{2-\mt{qh}}_{\mt{QH}} \, .
\end{split}
\end{equation}
At this point, a $\pi$-pulse flips the impurity spins, bringing the state $\ket{\Psi_2}$ to
\begin{equation}\label{supp:eq:psi_3}
\begin{split}
    \ket{\Psi_3} = & \frac{1}{2} \, \ket{\vc{r}_3 \downarrow,\vc{r}_4 \downarrow} 
    \hat{U}_{\vc{r}_1\to\vc{r}_3}\hat{U}_{\vc{r}_2\to\vc{r}_4}\ket{2-\mt{qh}}_{\mt{QH}} + \frac{1}{2} \, \ket{\vc{r}_3 \downarrow,\vc{r}_2 \uparrow} 
    \hat{U}_{\vc{r}_1\to\vc{r}_3}\ket{2-\mt{qh}}_{\mt{QH}} \\
    & + \frac{1}{2} \, \ket{\vc{r}_1 \uparrow,\vc{r}_4 \downarrow} 
    \hat{U}_{\vc{r}_2\to\vc{r}_4}\ket{2-\mt{qh}}_{\mt{QH}} + \frac{1}{2} \, \ket{\vc{r}_1 \uparrow,\vc{r}_2 \uparrow} 
    \ket{2-\mt{qh}}_{\mt{QH}} \, .
\end{split}
\end{equation}
In the next step, the $\ket{\uparrow}$-states are transferred from $(\vc{r}_1,\vc{r}_2)$ to $(\vc{r}_4,\vc{r}_3)$, leading to the state
\begin{equation}\label{supp:eq:psi_4}
\begin{split}
    \ket{\Psi_4} = & \frac{1}{2} \, \ket{\vc{r}_3 \downarrow,\vc{r}_4 \downarrow} 
    \hat{U}_{\vc{r}_1\to\vc{r}_3}\hat{U}_{\vc{r}_2\to\vc{r}_4}\ket{2-\mt{qh}}_{\mt{QH}} 
    + \frac{1}{2} \, \ket{\vc{r}_3 \downarrow,\vc{r}_3 \uparrow} 
    \hat{U}_{\vc{r}_1\to\vc{r}_3}\hat{U}_{\vc{r}_2\to\vc{r}_3}\ket{2-\mt{qh}}_{\mt{QH}} \\
    & + \frac{1}{2} \, \ket{\vc{r}_4 \uparrow,\vc{r}_4 \downarrow} 
    \hat{U}_{\vc{r}_1\to\vc{r}_4}\hat{U}_{\vc{r}_2\to\vc{r}_4}\ket{2-\mt{qh}}_{\mt{QH}} + \frac{1}{2} \, \ket{\vc{r}_4 \uparrow,\vc{r}_3 \uparrow} 
    \hat{U}_{\vc{r}_1\to\vc{r}_4}\hat{U}_{\vc{r}_2\to\vc{r}_3}\ket{2-\mt{qh}}_{\mt{QH}} \, .
\end{split}
\end{equation}
Finally, application of a $\pi/2-$pulse with a control phase $\phi_{c}$ to $\ket{\Psi_4}$ implements the transformations $\ket{\uparrow}\to \ket{+_{\phi_c}}=1/\sqrt{2} \big( \ket{\uparrow} + e^{i\phi_c} \ket{\downarrow} \big)$ and $\ket{\downarrow}\to \ket{-_{\phi_c}}= 1/\sqrt{2} \big( e^{-i\phi_c} \ket{\uparrow} - \ket{\downarrow} \big)$, resulting in the state $\ket{\Psi_f}$ given by
\begin{equation}\label{supp:eq:psif}
\begin{aligned}
    \ket{\Psi_{f}} 
    = & \frac{1}{2} \, \ket{\vc{r}_3 - _{\phi_c},\vc{r}_4 -_{\phi_c}} \hat{U}_{\vec{r}_1\to\vec{r}_3} \hat{U}_{\vc{r}_2\to\vc{r}_4}\ket{\rm 2-qh}_{\mathrm{QH}}
    \\ & + \frac{1}{2} \, \ket{\vec{r}_3 + _{\phi_c}, \vec{r}_4 + _{\phi_c}}
    \hat{U}_{\vec{r}_1\to\vec{r}_4} \hat{U}_{\vec{r}_2\to\vec{r}_3} \ket{\rm 2-qh}_{\mathrm{QH}}
    \\ & + \frac{1}{2} \, \ket{\vc{r}_3 - _{\phi_c},\vc{r}_3 + _{\phi_c}} 
    \hat{U}_{\vec{r}_1\to\vec{r}_3} \hat{U}_{\vec{r}_2\to\vec{r}_3} \ket{\rm 2-qh}_{\mathrm{QH}}
    \\ & + \frac{1}{2} \, \ket{\vc{r}_4 + _{\phi_c},\vc{r}_4 - _{\phi_c} } \hat{U}_{\vec{r}_1\to\vec{r}_4} \hat{U}_{\vec{r}_2\to\vec{r}_4} \ket{\rm 2-qh}_{\mathrm{QH}} \, .
\end{aligned}
\end{equation}
Measurement of the impurities in the joint state $\ket{\vc{r}_3 \uparrow,\vc{r}_4 \uparrow}$ only takes contribution from the first and last components in the right hand side of Eq.~\eqref{supp:eq:psif}, as the other components consist of impurities at the same position.
The unnormalized post-measurement state of the quantum Hall system $\ket{\Psi_{\mt{pm}}}=\big(\bra{\vc{r}_3\uparrow,\vc{r}_4\uparrow}\otimes\mathbb{I}_{\mt{QH}}\big)\ket{\Psi_f}$ is 
\begin{equation}
\begin{split}
    \ket{\Psi_{\mt{pm}}} & = \frac{1}{4} \, e^{-i2\phi_c} \, \hat{U}_{\vc{r}_1\to\vc{r}_3}\hat{U}_{\vc{r}_2\to\vc{r}_4}\ket{2-\mt{qh}}_{\mt{QH}} + \frac{1}{4} \, \hat{U}_{\vc{r}_1\to\vc{r}_4}\hat{U}_{\vc{r}_2\to\vc{r}_3}\ket{2-\mt{qh}}_{\mt{QH}} \, . 
\end{split}
\end{equation}
The post-measurement state $\ket{\Psi_{\mt{pm}}}$ can be cast into the form
\begin{equation}\label{supp:eq:psi_pm}
\begin{split}
    \ket{\Psi_{\mt{pm}}} & = \frac{1}{4} \, \hat{U}_{\vc{r}_1\to\vc{r}_4}\hat{U}_{\vc{r}_2\to\vc{r}_3}\Big( \mathbb{I} + e^{-i2\phi_c} \, \hat{U}_{\vc{r}_3\to\vc{r}_2}\hat{U}_{\vc{r}_4\to\vc{r}_1}\hat{U}_{\vc{r}_1\to\vc{r}_3}\hat{U}_{\vc{r}_2\to\vc{r}_4} \,\Big)\ket{2-\mt{qh}}_{\mt{QH}} ,\, 
\end{split}
\end{equation}
where in Eq.~\ref{supp:eq:psi_pm} we used the property $\hat{U}_{\vc{r}\to\vc{r}'}=\hat{U}^{-1}_{\vc{r}'\to\vc{r}}$. 
The second term in parentheses in Eq.~\eqref{supp:eq:psi_pm} acts on the two-quasihole state by translating the quasihole initially at $\vc{r}_2$ to $\vc{r}_1$ via $\vc{r}_3$, and the quasihole initially at $\vc{r}_1$ to $\vc{r}_2$ via $\vc{r}_4$, thereby exchanging the two quasiholes and contributing the exchange part of the total geometric phase. Simultaneously, each quasihole is transported along half of the closed path $\mathcal{C}\!\equiv\!\vc{r}_1\to\vc{r}_4\to\vc{r}_2\to\vc{r}_3\to\vc{r}_1$ in the presence of the background magnetic field. According to Eq.~\eqref{eq:UDef}], this yields an Aharonov-Bohm contribution to the wave function phase equal to that of a single quasihole traversing the full closed path:
\begin{equation}
 e^{*}/\hbar c \int_{\mathcal{C}_1} d\vc{x} \cdot \vc{\mathcal{A}}(\vc{x}) + e^{*}/\hbar c \int_{\mathcal{C}_2} d\vc{x} \cdot \vc{\mathcal{A}}(\vc{x}) = e^{*}/\hbar c \oint_{\mathcal{C}} d\vc{x} \cdot \vc{\mathcal{A}}(\vc{x}) \equiv \varphi_{\mathrm{AB}},
\end{equation}
where $C_{1,2}$ denote the paths (half of the closed path $\mathcal{C}$) followed by the two quasiholes, respectively.

Thus, for Abelian anyons the action of $\hat{U}_{\vc{r}_3\to\vc{r}_2}\hat{U}_{\vc{r}_4\to\vc{r}_1}\hat{U}_{\vc{r}_1\to\vc{r}_3}\hat{U}_{\vc{r}_2\to\vc{r}_4}$ in Eq.~\ref{supp:eq:psi_pm} acts as a phase factor $\mathrm{exp}[i\varphi_{\mathrm{geo}}]$ with $\varphi_{\mathrm{geo}} \!=\! \varphi_{\mathrm{AB}}+\varphi_{\mathrm{exc}}$ as in Eq.~\ref{eq:geometricPhase} of the main text. 
It follows that the norm of $\ket{\Psi_{\mt{pm}}}$ gives the measurement probability $p_{\uparrow\uparrow}=|1/4\big(1+\mt{exp}[i(\varphi_{\mt{geo}} -2 \phi_c)]\big)|^2=1/4\,\mt{cos}^2\big(\varphi_{\mt{geo}}/2 -\phi_c\big)$.

Note that the same reasoning holds for non-Abelian braiding, with $\hat{W}=\hat{U}_{\vc{r}_1\to\vc{r}_4}\hat{U}_{\vc{r}_4\to\vc{r}_2}\hat{U}_{\vc{r}_2\to\vc{r}_3}\hat{U}_{\vc{r}_3\to\vc{r}_1}$ where $\hat{W}$ is the Wilson loop operator in Eq.~\eqref{eq:W} of the main text.
The measurement probability now is $p_{\uparrow\uparrow}=1/16\,_{\mt{QH}}\bra{2-\mt{qh}}\big(2\mathbb{I}+ e^{-i2 \phi_c}\hat{W}+ e^{i2 \phi_c}\hat{W}^{\dagger}\big)\ket{2-\mt{qh}}_{\mt{QH}}=1/8\,\Big(1+|\langle\hat{W}\rangle|\mt{cos}(\varphi^{\psi} -2 \phi_c)\Big)$.

\section{Chern Insulator: Lowest Band Projection}
\begin{figure}
    \centering
    \includegraphics{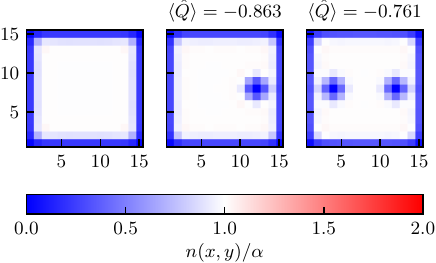}
    \caption{
        Projection to the lowest band does not change the results sizably, compare Fig.~\ref{fig:ChernInsulator:Density} of the main text.
        Density profiles for a system of $N=35$ particles in a magnetic field of flux $\alpha=0.2$ per plaquette subject to no (left), one (middle) and two (right) pinning potentials ($V_{\rm pin}/J = 1.5,~ \sigma/a=1.0$).
    }
    \label{supp:fig:ChernInsulator:Density}
\end{figure}

\begin{figure*}
    \centering
    \includegraphics{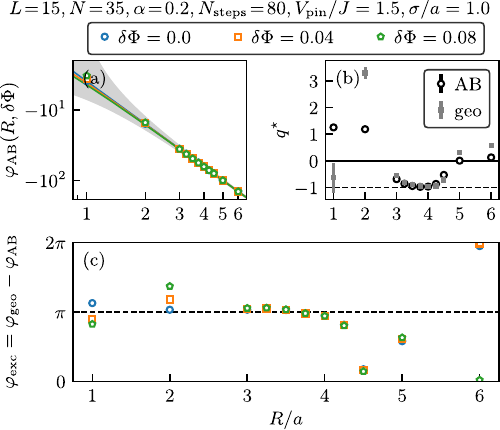}
    \caption{
        Projection to the lowest band does not change the results sizably, compare Fig.~\ref{fig:ChernInsulator:ABPhase:N35} of the main text.
        (a) Aharonov-Bohm phase for a single pinning potential ($V_{\rm pin}/J=1.5,~ \sigma/a=1$) moved along a circular path of radius $R/a$ in a system of $N=35$ particles on $15\times15$ sites with a flux $\alpha=0.2$ per plaquette, realizing a Chern insulator in the lowest Hofstadter band.
        The gray shaded area indicates a band of width $2\pi$ around the expected value for $\delta\Phi=0$.
        (b) Pinned charge $q^{\star}$ as extracted from the Aharonov-Bohm (circles) and full geometric (squares).
        For $3 \lesssim R/a \lesssim 4$ the pinned charge agrees well with the expected $q^{\star}\!=\!-1$.
        (c) Extracted exchange phase for two pinning potentials.
        For $3 \lesssim R/a \lesssim 4$, the phase is consistent with the expected $\varphi_{\rm exc}=\pi$, whereas for smaller radii the effect of not fully encircling the additional flux $\delta\Phi$ is visible for $\delta\Phi\neq0$.
        For larger radii, edge effects become substantial and lead to deviations from the expected exchange phase.
    }
    \label{supp:fig:ChernInsulator:ABPhase:Projected}
\end{figure*}

We perform numerical exchange experiments on non-interacting Chern insulators projected to the lowest Chern band.
These simulations might in principle yield slightly more robust results as the localized pinning potentials can lead to band mixing effects.
However, for our system of $N=35$ fermions on $15\times15$ sites we find no sizable changes to the density profile (lower row in Fig.~\ref{supp:fig:ChernInsulator:Density}) or the extracted Aharonov-Bohm or exchange phases (Fig.~\ref{supp:fig:ChernInsulator:ABPhase:Projected}).

\end{document}